\documentclass[oldversion]{aa}
\usepackage{graphicx}
\usepackage{txfonts}
\usepackage{natbib}
\usepackage{amssymb}
\newfont{\gwpfont}{cmssq8 scaled 1000}
\newcommand{\rexcess}{{\gwpfont REXCESS}}

\begin{document}
%
\def\etal{et al.}

\def\Mgv{M_{\rm g,500}}
\def\Mg{M_{\rm g}}
\def\YX {Y_{\rm X}}
\def\TX {T_{\rm X}}
\def\fgv {f_{\rm g,500}}
\def\fg  {f_{\rm g}}
\def\kT {{\rm k}T}
\def\Mv {M_{\rm 500}}
\def \Rv {R_{500}}
\def\keV {\rm keV}

\def\MT {$M_{500}$--$T_{\rm X}$}
\def\MY {$M_{500}$--$Y_{\rm X}$}
\def\MMg {$M_{500}$--$M_{\rm g,500}$}
\def\MgT {$M_{\rm g,500}$--$T_{\rm X}$}
\def\MgY {$M_{\rm g,500}$--$Y_{\rm X}$}
\def\LxT {$L$--$T$}
\def\LxYx {$L-Y_X$}
\def\YxM {$Y_X-\Mv$}
\def\LxM {$L-\Mv$}
\def\fgas {$f_{\rm gas}$}

\def\msol {{\rm M_{\odot}}}

\def\lesssim{\mathrel{\hbox{\rlap{\hbox{\lower4pt\hbox{$\sim$}}}\hbox{$<$}}}}
\def\gtrsim{\mathrel{\hbox{\rlap{\hbox{\lower4pt\hbox{$\sim$}}}\hbox{$>$}}}}

\newcommand{\propsim}{\lower 3pt \hbox{$\, \buildrel {\textstyle
       \propto}\over {\textstyle \sim}\,$}}

\def \xmm {\hbox{\it XMM-Newton}}
\def \chandra {\hbox{\it Chandra }}

\title{Gas entropy in a representative sample of nearby X-ray galaxy clusters (REXCESS): relationship to gas mass fraction}   
\author{G.W. Pratt\inst{1,2}, M. Arnaud\inst{1}, R. Piffaretti\inst{1}, H. B\"ohringer\inst{2}, T.J. Ponman\inst{3}, J.H. Croston\inst{4}, G.M. Voit\inst{5}, S. Borgani\inst{6} and~R.G.~Bower\inst{7}
}
\offprints{G.W. Pratt, \email{gabriel.pratt@cea.fr}}
\authorrunning{G.W. Pratt et al.}
\titlerunning{Entropy properties of the \rexcess}
 \institute{
 $^1$ Laboratoire AIM, IRFU/Service d'Astrophysique - CEA/DSM - CNRS - Universit\'{e} Paris Diderot, B\^{a}t. 709, CEA-Saclay, F-91191 Gif-sur-Yvette Cedex, France \\ 
 $^2$ Max-Planck-Institut f\"ur extraterrestriche Physik, Giessenbachstra{\ss}e, 85748 Garching, Germany \\
$^3$ School of Physics and Astronomy, University of Birmingham, Edgbaston, Birmingham B15 2TT, UK\\
$^4$ School of Physics and Astronomy, University of Southampton, Southampton, Hampshire,  SO17 1BJ, UK  \\
$^5$ Department of Physics and Astronomy, Michigan State University, East Lansing, MI 48824-2320, USA \\
$^6$ Dipartimento  di Astronomia dell'Universit\`{a} di Trieste, via Tiepolo 11, 34131 Trieste, Italy \\
$^7$ Institute for Computational Cosmology, Department of Physics, Durham University, South Road, Durham DH1 3LE, UK
}

  \date{Received 17 September 2009; accepted 10 November 2009}
  \abstract 
   {We examine the radial entropy distribution and its scaling using 31 nearby galaxy clusters from the Representative \xmm\ Cluster Structure Survey (\rexcess), a sample in the temperature range 2-9 keV selected in X-ray luminosity only, with no bias toward any particular morphological type. The entropy profiles are robustly measured at least out to $R_{1000}$ in all systems and out to $\Rv$ in thirteen systems. Compared to theoretical expectations from  non-radiative cosmological simulations, the observed distributions show a radial and mass-dependent excess entropy, such that the excess is greater and extends to larger radii in lower mass systems. At $\Rv$, the mass dependence and entropy excess are both negligible within the large observational and theoretical uncertainties. Mirroring this behaviour, the scaling of gas entropy is shallower than self-similar in the inner regions, but steepens with radius, becoming consistent with self-similar at $\Rv$. There is a large dispersion in scaled entropy in the inner regions, apparently linked to the presence of cool cores and dynamical activity; at larger radii the dispersion decreases by approximately a factor of two to 30 per cent, and the dichotomy between subsamples disappears. There are two peaks in the distribution of both inner slope and, after parameterising the profiles with a power law plus constant model, in  central entropy $K_0$. However, we are unable to distinguish between a bimodal or a left-skewed distribution of $K_0$ with the present data.  The distribution of outer slopes is unimodal with a median value of $0.98$, and there is a clear correlation of outer slope with temperature. Renormalising the dimensionless entropy profiles by the gas mass fraction profile $f_{\rm gas} (< R)$, leads to a remarkable reduction in the scatter, implying that gas mass fraction variations with radius and mass are the cause of the observed entropy structural and scaling properties. The results are consistent with the picture of a cluster population in which entropy modification is centrally concentrated and extends to larger radii at lower mass, leading to both a radial and a mass-dependence in the gas mass fraction, but which is increasingly self-similar at large radius. The observed normalisation, however, would suggest entropy modification at least up to $R_{1000}$, and even beyond, in all but the most massive systems. We discuss a tentative scenario to explain the observed behaviour of the entropy and gas mass fraction in the \rexcess\ sample, in which  a combination of extra heating and merger mixing maintains an elevated central entropy level in the majority of the population, and a smaller fraction of systems is able to develop a cool core.
}

   \maketitle
%

\section{Introduction}

\begin{table*}[]
\begin{minipage}{\textwidth}
\begin{center}
\caption{{\footnotesize Basic cluster data. $K(R_\delta)$ is the entropy measured at the radius corresponding to density contrast $\delta$. Cool core and morphological classification from \citet{pratt09}.\label{tab:Kscaled}}}
\centering
{\tiny
\begin{tabular}{l r r r r r r r r r}
\hline
\hline
\\
\multicolumn{1}{c}{Cluster} & \multicolumn{1}{c}{$z$} & \multicolumn{1}{c}{$kT$}\footnote{Temperature  in keV, measured in the $[0.15-0.75]\,\Rv$ aperture} & \multicolumn{1}{c}{$M_{500}$}\footnote{Mass in units of $h_{70}^{-1}\, 10^{14}\,M_\odot$, estimated from the $M_{500}-Y_X$ relation given in Eqn.~\ref{eqn:Yx} \citep[][]{arnaud09}.} & \multicolumn{1}{c}{$K\, (0.1\,R_{200})$} & \multicolumn{1}{c}{$K\, (R_{2500})$} & \multicolumn{1}{c}{$K\, (R_{1000})$} & \multicolumn{1}{c}{$K\, (R_{500})$} & \multicolumn{1}{c}{CC} & \multicolumn{1}{c}{Disturbed}\\
\hline
\\
RXC\,J0003.8+0203 & 0.0924 & $3.83_{-0.11}^{+0.11}$ & $2.11_{-0.04}^{+0.04}$ & $   273.35 \pm  17.67 $ & $   619.88 \pm  50.40 $ & $   913.29 \pm 114.31 $ & \ldots  & \ldots & \ldots \\

RXC\,J0006.0-3443 & 0.1147 & $5.24_{-0.21}^{+0.21}$ & $3.95_{-0.12}^{+0.12}$ & $   469.36 \pm  55.82 $ & $   786.71 \pm  80.54 $ & $  1040.90 \pm  98.14 $ & \ldots  & \ldots & \checkmark \\

RXC\,J0020.7-2542 & 0.1410 & $5.54_{-0.13}^{+0.13}$ & $3.84_{-0.06}^{+0.06}$ & $   347.10 \pm  28.49 $ & $   727.97 \pm  58.46 $ & $  1085.23 \pm 143.63 $ & $  1498.66 \pm 687.71 $ & \ldots & \ldots \\

RXC\,J0049.4-2931 & 0.1084 & $2.87_{-0.10}^{+0.10}$ & $1.62_{-0.04}^{+0.04}$ & $   191.06 \pm  14.02 $ & $   442.17 \pm  74.37 $ & $   647.11 \pm  82.13 $ &  \ldots   & \ldots & \ldots \\

RXC\,J0145.0-5300 & 0.1168 & $5.81_{-0.15}^{+0.15}$ & $4.37_{-0.08}^{+0.08}$ & $   365.72 \pm  26.31 $ & $   840.65 \pm  58.67 $ & $  1400.17 \pm 166.17 $ & \ldots & \ldots & \checkmark \\

RXC\,J0211.4-4017 & 0.1008 & $2.08_{-0.05}^{+0.05}$ & $1.00_{-0.02}^{+0.02}$ & $   141.17 \pm   8.83 $ & $   309.39 \pm  22.19 $ & $   522.67 \pm  47.23 $ & $   687.20 \pm 134.45 $ & \ldots & \ldots \\

RXC\,J0225.1-2928 & 0.0604 & $2.53_{-0.14}^{+0.14}$ & $0.96_{-0.04}^{+0.04}$ & $   173.95 \pm  13.33 $ & $   546.12 \pm 114.66 $ & $   747.10 \pm  81.97 $ &  \ldots  & \ldots & \checkmark \\

RXC\,J0345.7-4112 & 0.0603 & $2.28_{-0.06}^{+0.07}$ & $0.97_{-0.02}^{+0.02}$ & $   175.30 \pm  13.91 $ & $   384.26 \pm  38.58 $ & $   481.23 \pm  40.90 $ & \ldots  & \checkmark & \ldots \\

RXC\,J0547.6-3152 & 0.1483 & $6.04_{-0.14}^{+0.14}$ & $4.98_{-0.08}^{+0.09}$ & $   324.96 \pm  16.61 $ & $   842.16 \pm  61.88 $ & $  1128.97 \pm  96.51 $ &  \ldots & \ldots & \ldots  \\

RXC\,J0605.8-3518 & 0.1392 & $4.93_{-0.12}^{+0.12}$ & $3.87_{-0.06}^{+0.06}$ & $   238.61 \pm  13.12 $ & $   631.88 \pm  32.69 $ & $  1167.71 \pm 157.17 $ & $  1745.33 \pm 462.31 $ & \checkmark & \ldots \\

RXC\,J0616.8-4748 & 0.1164 & $4.18_{-0.11}^{+0.11}$ & $2.70_{-0.05}^{+0.06}$ & $   348.61 \pm  39.07 $ & $   639.08 \pm  40.11 $ & $   939.62 \pm  77.66 $ & $  1357.27 \pm 224.37 $ & \ldots & \checkmark \\

RXC\,J0645.4-5413 & 0.1644 & $7.23_{-0.18}^{+0.18}$ & $7.38_{-0.14}^{+0.14}$ & $   349.48 \pm  27.52 $ & $   941.30 \pm  79.96 $ & $  1462.99 \pm 158.42 $ & \ldots & \ldots & \ldots \\

RXC\,J0821.8+0112 & 0.0822 & $2.81_{-0.11}^{+0.10}$ & $1.31_{-0.04}^{+0.03}$ & $   285.87 \pm  19.08 $ & $   436.36 \pm  45.57 $ & $   617.55 \pm  56.35 $ & \ldots  & \ldots & \ldots \\

RXC\,J0958.3-1103 & 0.1669 & $5.95_{-0.33}^{+0.49}$ & $4.17_{-0.15}^{+0.22}$ & $   220.23 \pm  33.22 $ & $   875.57 \pm 297.68 $ & $  1421.19 \pm 618.50 $ & \ldots & \checkmark & \ldots \\ 

RXC\,J1044.5-0704 & 0.1342 & $3.58_{-0.05}^{+0.05}$ & $2.27_{-0.02}^{+0.02}$ & $   164.67 \pm   6.09 $ & $   447.65 \pm  21.65 $ & $   722.02 \pm  84.78 $ & $  1021.75 \pm 127.41 $ & \checkmark & \ldots  \\

RXC\,J1141.4-1216  & 0.1195 & $3.58_{-0.06}^{+0.06}$ & $2.27_{-0.02}^{+0.02}$ & $   197.66 \pm   7.26 $ & $   557.71 \pm  25.24 $ & $   849.48 \pm  74.07 $ & $  1016.40 \pm  91.66 $ & \checkmark & \ldots  \\

RXC\,J1236.7-3354 & 0.0796 & $2.77_{-0.05}^{+0.06}$ & $1.33_{-0.02}^{+0.02}$ & $   179.13 \pm   6.56 $ & $   469.68 \pm  25.36 $ & $   849.80 \pm 133.02 $ & \ldots & \ldots & \ldots \\

RXC\,J1302.8-0230 & 0.0847 & $3.48_{-0.08}^{+0.08}$ & $1.89_{-0.03}^{+0.03}$ & $   310.02 \pm  16.16 $ & $   741.95 \pm  43.99 $ & $   779.02 \pm  89.95 $ & $   684.30 \pm  52.30 $ & \checkmark & \checkmark \\

RXC\,J1311.4-0120 & 0.1832 & $8.67_{-0.12}^{+0.12}$ & $8.41_{-0.08}^{+0.08}$ & $   337.47 \pm  10.26 $ & $  1044.38 \pm  56.40 $ & $  1653.59 \pm 116.17 $ & \ldots & \checkmark & \ldots \\

RXC\,J1516.3+0005 & 0.1181 & $4.68_{-0.08}^{+0.13}$ & $3.28_{-0.04}^{+0.07}$ & $   321.83 \pm  16.45 $ & $   673.65 \pm  35.14 $ & $  1004.69 \pm  96.87 $ & \ldots & \ldots & \ldots \\

RXC\,J1516.5-0056 & 0.1198 & $3.70_{-0.08}^{+0.10}$ & $2.59_{-0.04}^{+0.05}$ & $   314.75 \pm  22.64 $ & $   712.59 \pm  51.51 $ & $   728.24 \pm  54.37 $ & $   732.28 \pm  50.80 $ & \ldots & \checkmark \\

RXC\,J2014.8-2430 & 0.1538 & $5.75_{-0.10}^{+0.10}$ & $5.38_{-0.07}^{+0.07}$ & $   263.67 \pm  12.80 $ & $   726.64 \pm  39.70 $ & $  1158.45 \pm  96.11 $ & \ldots & \checkmark & \ldots \\

RXC\,J2023.0-2056 & 0.0564 & $2.72_{-0.09}^{+0.09}$ & $1.21_{-0.03}^{+0.03}$ & $   248.64 \pm  15.10 $ & $   527.38 \pm  58.61 $ & $   657.84 \pm  42.38 $ & \ldots & \ldots & \checkmark \\

RXC\,J2048.1-1750 & 0.1475 & $5.06_{-0.11}^{+0.11}$ & $4.32_{-0.07}^{+0.07}$ & $   409.74 \pm  35.68 $ & $   743.70 \pm  46.76 $ & $  1006.16 \pm  53.29 $ & $  1086.97 \pm  79.89 $ & \ldots & \checkmark \\

RXC\,J2129.8-5048 & 0.0796 & $3.84_{-0.14}^{+0.14}$ & $2.26_{-0.06}^{+0.06}$ & $   441.52 \pm  31.94 $ & $   644.40 \pm  53.66 $ & $   861.48 \pm  67.50 $ & \ldots & \ldots & \checkmark \\

RXC\,J2149.1-3041 & 0.1184 & $3.48_{-0.07}^{+0.07}$ & $2.25_{-0.03}^{+0.03}$ & $   196.26 \pm   9.85 $ & $   513.99 \pm  32.71 $ & $   762.82 \pm  80.24 $ & $  1062.45 \pm 185.45 $ & \checkmark & \ldots \\

RXC\,J2157.4-0747 & 0.0579 & $2.79_{-0.07}^{+0.07}$ & $1.29_{-0.03}^{+0.03}$ & $   390.50 \pm  40.35 $ & $   659.44 \pm  65.91 $ & $   675.83 \pm  23.62 $ & \ldots & \ldots & \checkmark \\

RXC\,J2217.7-3543 & 0.1486 & $4.62_{-0.08}^{+0.09}$ & $3.61_{-0.05}^{+0.05}$ & $   282.04 \pm  13.59 $ & $   610.69 \pm  34.87 $ & $   872.97 \pm  62.41 $ & $  1076.81 \pm 107.72 $ & \ldots & \ldots \\

RXC\,J2218.6-3853 & 0.1411 & $6.18_{-0.20}^{+0.20}$ & $4.92_{-0.11}^{+0.11}$ & $   263.03 \pm  12.24 $ & $   859.14 \pm  64.35 $ & $  1610.20 \pm 225.47 $ & $  2278.44 \pm 862.21 $ & \ldots & \checkmark \\

RXC\,J2234.5-3744 & 0.1510 & $7.32_{-0.12}^{+0.12}$ & $7.36_{-0.09}^{+0.09}$ & $   451.39 \pm  28.73 $ & $   813.38 \pm  42.42 $ & $  1170.94 \pm  97.12 $ & \ldots  & \ldots & \ldots \\

RXC\,J2319.6-7313 & 0.0984 & $2.56_{-0.08}^{+0.08}$ & $1.56_{-0.03}^{+0.03}$ & $   133.95 \pm   7.23 $ & $   380.37 \pm  32.90 $ & $   637.12 \pm  66.10 $ & $  1017.21 \pm 211.39 $ &\checkmark & \checkmark \\
\\
\hline
\end{tabular}
}
\end{center}
\end{minipage}
\end{table*}

The first order model of structure formation -- that of hierarchical, dark matter dominated gravitational collapse -- is capable of reproducing only the gross statistical properties of the galaxy cluster population. In this scenario, the intracluster medium (ICM) is heated to X-ray emitting temperatures by shocks and compression as it falls into the potential well of the dark matter, and the resulting X-ray cluster population is self-similar and scale-free. In real clusters, second order effects, linked primarily to feedback from galaxy formation and radiative cooling of the gas, serve to modify the X-ray properties of the ICM with respect to these expectations \citep[see, e.g.,][for recent reviews]{voit05,bk09}. The effect of these nongravitational processes is substantial. It can be seen most readily in the relation between the X-ray luminosity and temperature, which in the first order scenario scales simply as $L \propsim T^2$, but which is observed to scale as $L \propsim T^3$ \citep[e.g.][and references therein]{pratt09}, implying a progressive suppression of luminosity in low temperature systems.

In recent years, spatially resolved observations have allowed us to examine in more detail the impact of nongravitational processes on the ICM, mainly through radial profiles and mapping. In this context, the entropy $K$ of the ICM\footnote{Keeping with convention, we use the X-ray astronomer's 'entropy' throughout this paper. Defined as $K = kT/n_e^{2/3}$, where $n_e$ is the electron number density, this quantity is related to the true entropy by a logarithm and an additive constant.} is of considerable interest because the observable X-ray characteristics of a cluster are just manifestations of its distribution in the dark matter potential well. Entropy is generated during the hierarchical assembly process, yet is modified by any other process that can change the physical characteristics of the gas. It is thus a quantity that preserves a record both of the accretion history of a cluster and of the influence of non-gravitational processes on the properties of its ICM, and as such it is a useful tool for our understanding of the thermodynamic history of the cluster population. 

Early measurements of the entropy based primarily on {\it ROSAT} and {\it ASCA} data indicated that groups had flatter entropy profiles than cluster scale objects \citep{djf96}, and measurements of the entropy at $0.1\,R_{200}$ revealed an entropy-temperature ($K-T$) relation that was shallower than expected \citep{pcn99,lpc00,psf03}. These data also afforded the first indications for excess entropy above that expected from gravitational collapse even at large radius in group-scale objects \citep{fin02,psf03}. The advent of \xmm\ and {\it Chandra} has allowed relatively high resolution spatially resolved measurement of the entropy to be obtained across a wide range of cluster and group masses \citep{pa03,piff05,pap06,me07,nkv07,zhang08,sanderson09,johnson09}. Recent results have suggested that the entropy is indeed higher than expected from gravitational collapse at least out to $R_{2500}$ \citep{pap06,nkv07,sun09}, and perhaps further \citep{pap06,sun09}, even up to relatively high masses. In addition, indications for excess entropy have been found at large radius in intermediate redshift groups \citep{jeltema06}, and the first measurements of the entropy evolution have been undertaken \citep{ettori04}.  

In the present paper we re-investigate the entropy with \rexcess\ \citep{boehringer07}, a representative sample of 33 local ($z < 0.2$) clusters drawn from the REFLEX catalogue \citep{reflex}, all of which have been observed with \xmm. The properties of the \rexcess\ sample allow us to define a robust local reference for entropy structure and scaling. \rexcess\ was designed to be representative of any high-quality local X-ray survey, thus clusters have been selected in luminosity only, ensuring no morphological bias, in such a way as to sample the X-ray cluster luminosity function in an optimal manner. Moreover, distances were optimised so that $R_{500}$ falls well within the \xmm\ field of view, increasing the precision of measurements at large radii as compared to more nearby clusters, which often fill the field of view and for which background modelling is consequently more complicated.

In the following, we first examine the normalisation of the entropy with respect to predictions from non-radiative cosmological simulations  -- such `adiabatic' simulations include only gravitational processes -- finding a systematic entropy excess that is greater at small radii and in lower mass systems. The considerable dispersion at small radii appears linked to whether a cluster possesses a cool core or is morphologically disturbed. The mass dependence disappears at $\Rv$, implying that entropy scaling is self-similar, with a normalisation that is approximately consistent with predictions. Parameterising the profiles in terms of a power law plus constant model, there are two peaks in the distribution of central entropy but there is no strong evidence that it is bimodal. The distribution of outer slopes is unimodal and the slope depends on temperature. We then link the entropy scaling and structural properties to a systematic variation in gas content with total mass and with radius. Finally, we discuss mechanisms which could bring about the observed entropy characteristics, and propose a tentative scenario to explain the observed entropy distributions.

We adopt a $\Lambda$CDM cosmology with $H_0= 70$ km s$^{-1}$ Mpc$^{-1}$ (i.e., $h_{70} = 1$), $\Omega_M=0.3$ and $\Omega_\Lambda=0.7$.
All uncertainties are quoted at the 68 per cent confidence level.


\section{Sample and analysis}

\subsection{Sample description and subsample definition}
\label{sec:sample}

A full description of the \rexcess\ sample, including the \xmm\ observation details, can be found in \citet{boehringer07}, and the preliminary data analysis is described in \citet{croston08}. Two of the objects, RCXC\,J0956.4-1004 and RXC\,J2157.4-0747 (the Abell 901/902 supercluster and a bimodal cluster, respectively), display complex morphologies which preclude their use for the present radial profile analysis. Basic cluster parameters are listed in Table~\ref{tab:Kscaled}.

On occasion in the following, we will subdivide the sample into cool core and non-cool core systems, or according to whether the clusters are morphologically relaxed or unrelaxed. These subsamples were established to cull approximately the most extreme thirty percent of the full sample in each category and are defined as in \citet{pratt09}. Thus clusters with central density $E(z)^{-2} n_{e,0} > 4 \times 10^{-2}$ cm$^{-3}$ are classified\footnote{$E(z)$ is the ratio of the Hubble constant at redshift z to its present value, $E^{2}(z) = \Omega_{\rm M}(1 + z)^3 + \Omega_\Lambda$.}  as cool core systems (10/31), and those with centre shift parameter $\langle w \rangle > 0.01\,\Rv$ (derived with the central regions excised) are classified as morphologically disturbed (12/31). Both the central densities $n_{e,0}$ and centroid shift parameter $\langle w \rangle$ are given in \citet{haarsma09}.


\subsection{Data analysis}

Event lists were processed using version 7.0 of the \xmm\ SAS. All data products were extracted from event lists that were generated, cleaned, {\sc pattern}-selected, vignetting-corrected, and point source-removed as described in \citet{pratt07}.

\subsubsection{Gas density profiles}

The procedure used to calculate the gas density profiles, plus extensive analysis of their properties, is described in full in \citet{croston08}. In brief, surface brightness profiles, centred on the peak of the X-ray emission, were extracted from $3\farcs3$ bins in the [0.3-2] keV band and deprojected and PSF corrected using the non-parametric method introduced in \citet{croston06}. These were converted to gas density by calculating a conversion factor in XSPEC using the temperature in the $[0.15-1]\,\Rv$ aperture, and subsequently corrected to take into account radial variations of temperature and abundance to give the final deprojected, PSF-corrected radial density profile. 


\subsubsection{Temperature profiles}

Projected 2D temperature profiles were derived from spectra extracted in logarithmically-spaced annular bins centred on the peak of the X-ray emission.  Binning was such that the first bin was defined to have a significance of $30\sigma$ above background, and subsequent bins were defined so as to have $R_{\rm out}/R_{\rm in} = 1.33-1.5$ depending on the quality of the observation. The instrumental and particle background was subtracted from each annulus using custom stacked, recast data files accumulated from observations obtained with the filter wheel in the CLOSED position (FWC), renormalised using the count rate in a high energy band free of cluster emission\footnote{Our adoption of FWC data allows the use of a physical model for the X-ray background, in contrast to our previous analysis which used a blank sky background \citep{pratt07}. The results are consistent.}. After subtraction of the FWC spectra, all spectra were grouped to a minimum of 25 counts per bin. 

\begin{figure*}[ht]
\begin{centering}
\includegraphics[scale=1.,angle=0,keepaspectratio,width=1.05\columnwidth]{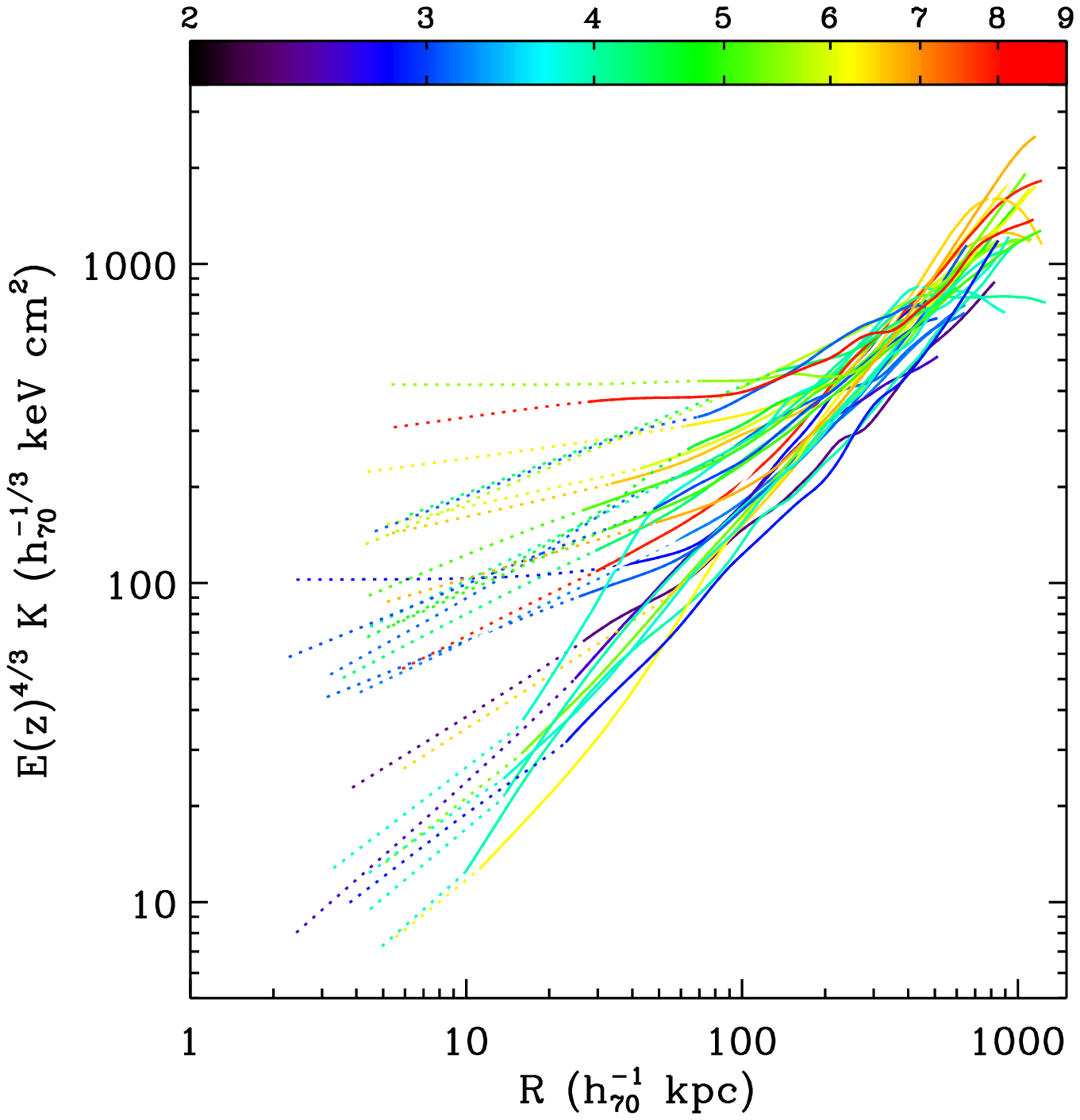}
\hfill
\includegraphics[scale=1.,angle=0,keepaspectratio,width=1.05\columnwidth]{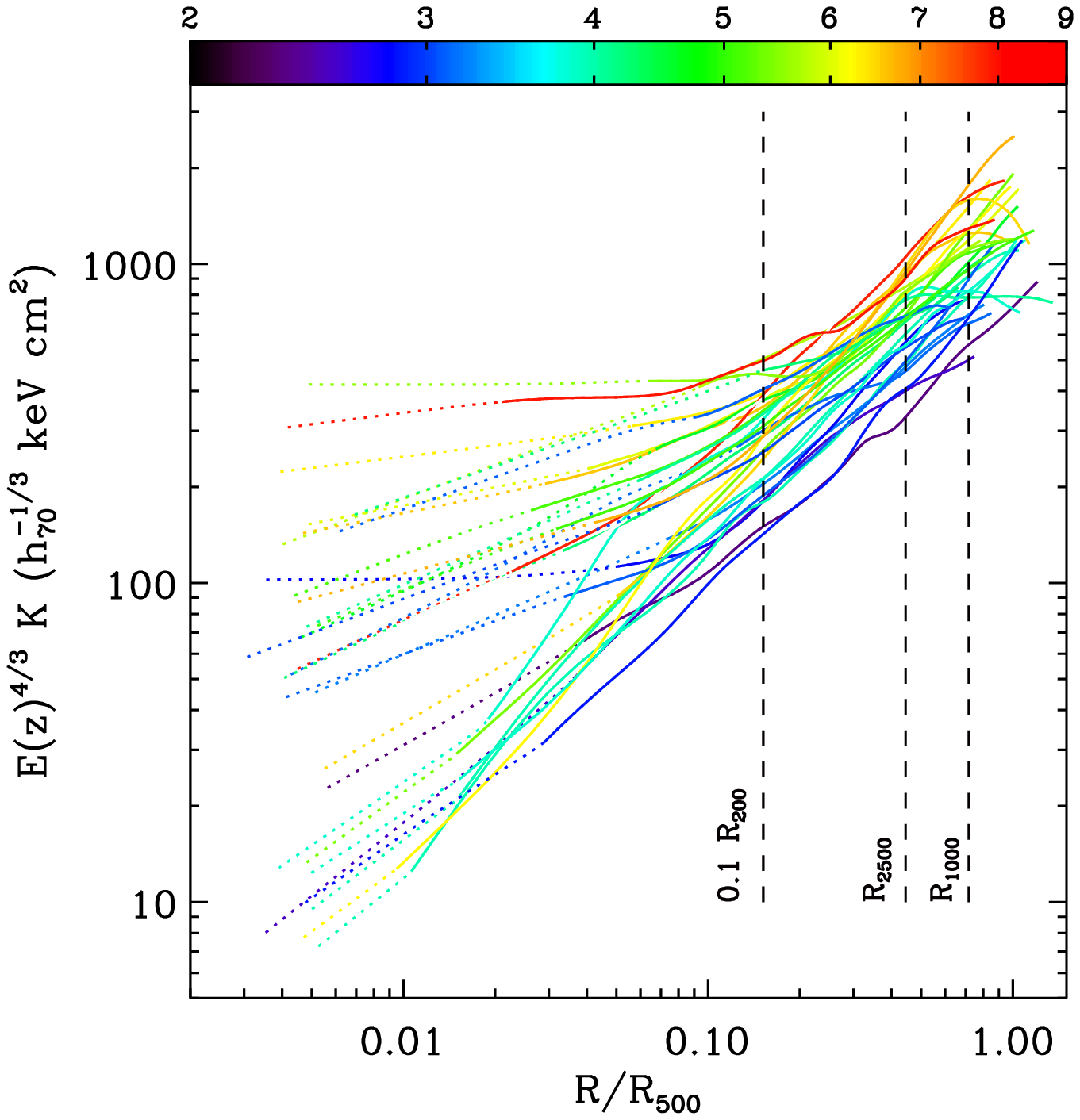}
\end{centering}
\caption{{\footnotesize Entropy profiles of the \rexcess\ sample, colour coded according to spectroscopic temperature measured in the $[0.15-0.75]\,\Rv$ aperture. Solid lines show the profiles derived from direct measurement; dotted lines show the entropy extrapolated into the central regions assuming an isothermal  distribution at the temperature of the inner 3D data point (see text). On the left, the profiles are plotted in physical units; on the right, they are plotted in units of scaled radius $\Rv$ estimated from the $M_{500}$--$Y_X$ relation given in Eqn.~\ref{eqn:Yx}. In the right hand plot, dashed lines mark, from left to right, radii corresponding to $0.1\,R_{200}$, $R_{2500} (\sim 0.45\,R_{500})$ and $R_{1000} (\sim 0.7\,R_{500})$. }}\label{fig:Kprof}
\end{figure*}

We then fitted the FWC-subtracted spectrum of an annular region external to the cluster emission with a physically motivated model of the cosmic X-ray background, consisting of two {\sc MeKaL} components plus an absorbed power law with a fixed slope of $\Gamma=1.4$ \citep[see][]{lumb02, deluca04}. Spectra were fitted in the $[0.3 - 10]$ keV range using $\chi^2$ statistics, excluding the [1.4-1.6] keV band (due to the Al line in all three detectors), and, in the EPN, the [7.45-9.0] keV band (due to the strong Cu line complex). In these fits the {\sc MeKaL} models are unabsorbed and have solar abundances, and the temperature and normalisations are free parameters; the powerlaw component is absorbed by the Galactic absorption in the direction of the cluster and since it has a fixed slope, only its normalisation is an additional free parameter in the fit. This best fitting background model, with renormalisation appropriate to the ratio of the surface area of the extraction regions (corrected for chip gaps, etc.), is then added as an extra component in each annular fit. This is our standard fit. 

We then vary the power law index in such a way as to mimic a $\pm 10$ per cent variation in the [2-10] keV flux and refit the spectrum of the external region. The annular spectra were refitted using this new cosmic X-ray background model, and the change in the cluster temperature in each annulus relative to the standard fit was treated as a systematic uncertainty and added in quadrature to the statistical errors in each annulus.

To deproject and PSF-correct, we assume that the 3D temperature profile can be represented by a parametric model \citep[adapted from][]{vikh06} that is convolved with a response matrix which simultaneously takes into account projection and PSF redistribution. 
This model was projected, taking into account the weighting scheme proposed by \citet[][see also \citealt{mazz04}]{vikhw} to correct for the bias introduced by fitting isothermal models to multi-temperature plasma emission, and fitted to the observed 2D annular temperature profile. Uncertainties were estimated from Monte Carlo randomisation of the projected temperature profile assuming a Gaussian distribution defined by the uncertainties on each data point, and then corrected to take into account the fact that parametric models tend to over-constrain the 3D profile. Full details of the deprojection and PSF correction of the temperature profiles, plus extensive tests of the robustness of the method, will be detailed in a forthcoming paper.


\subsubsection{Entropy profiles}

Since the density profiles are determined on a radial grid of significantly higher resolution than that of the temperature profiles, we determined the best fitting parametric 3D temperature profile on the same grid as that of the deprojected, deconvolved density profile and calculated the entropy, $K = kT/n_e^{2/3}$, accordingly. 

In all cases, in the very central regions a single temperature bin encompasses a region covered by several density profile bins (the median number is 5). Given that the central density of the galaxy cluster population exhibits a dispersion of up to two orders of magnitude and the overall density profile changes by up to three orders of magnitude from the centre to the outskirts \citep[e.g.,][]{croston08}, while the temperature varies only by a factor of 2-3 \citep[e.g.,][]{pratt07}, it is clear that the characteristics of the density drive the properties of the entropy profiles. In order to examine the behaviour of the central entropy, we assume a constant central temperature, with the value given by the 3D temperature of the first shell. A similar procedure was used by \citet[][]{donahue05} and \citet{cav09} and our adoption of this approach allows us to compare directly with their results. Note that for systems with poor central temperature profile resolution this extrapolation is only weakly model dependent, since it essentially concerns the disturbed systems, which have rather flat central temperature profiles \citep[see Figure 3 of][]{arnaud09}. When this scheme is applied, the \xmm\ profiles have a typical central resolution of $\sim 5\,h_{70}^{-1}$ kpc, which compares favourably with the typical resolution of $\sim 2$ kpc in the {\it Chandra\/} analysis of \citet{cav09}. The left hand panel of Figure~\ref{fig:Kprof} shows these entropy profiles plotted in physical units ($h_{70}^{-1}$ kpc). 


\begin{figure*}[]
\begin{centering}
\includegraphics[scale=1.,angle=0,keepaspectratio,width=1.05\columnwidth]{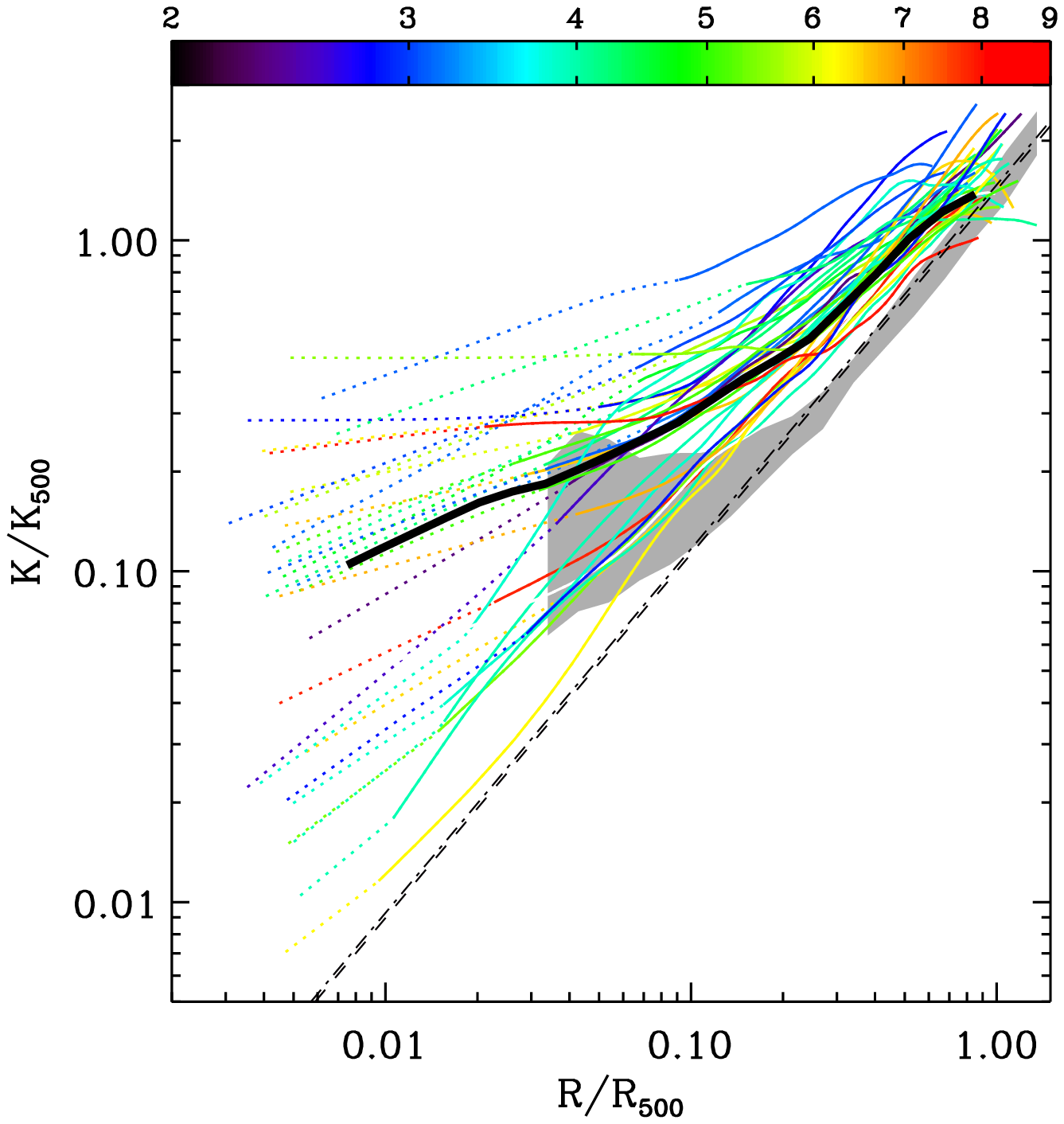}
\hfill
\includegraphics[scale=1.,angle=0,keepaspectratio,width=1.05\columnwidth]{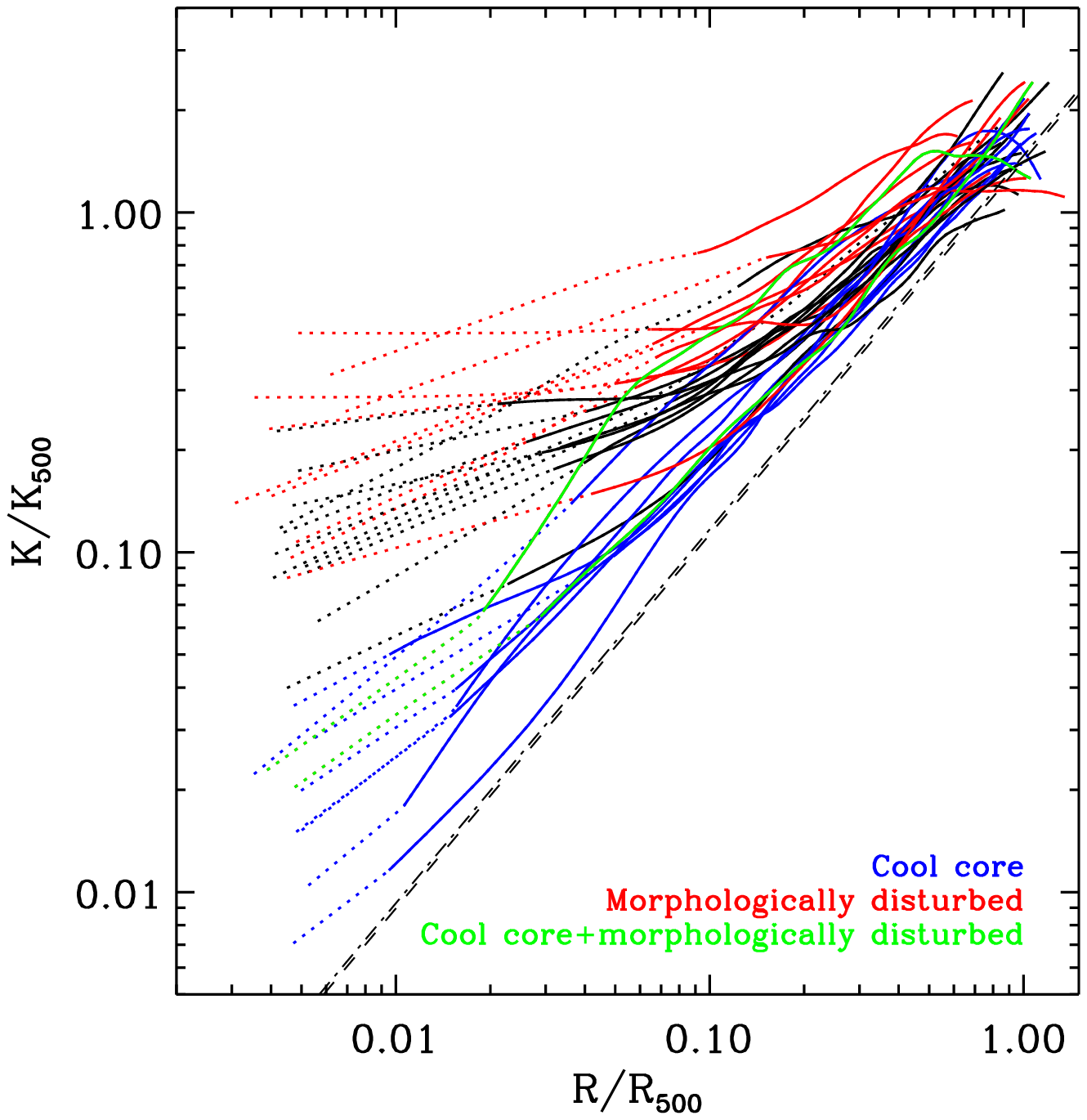}
\end{centering}
\caption{{\footnotesize Dimensionless entropy profiles of the \rexcess\ sample compared to theoretical expectations from non-radiative simulations. The observed profiles have been renormalised by the characteristic $K_{500}$ as defined in Equation~\ref{eqn:K500}. Line styles are as for Figure~\ref{fig:Kprof}. The dashed line depicts the best fitting power law fit to the the median entropy profile in the radial range $[0.1 - 1] R_{200}$ for the clusters formed in the non-radiative simulations of \citet{vkb05}. The dot-dashed line represents the same relation corrected for a 13 per cent underestimate of true mass due to the use of hydrostatic equilibrium. {\it Left panel:} clusters arranged according to temperature. The thick black line represents the median of all observed dimensionless profiles. The shaded grey area corresponds to the region enclosed by the median profile and typical scatter of the SPH simulations in \citet{vkb05}. {\it Right panel:} subsamples identified, defined as in Section~\ref{sec:sample}. Black profiles denote clusters that are neither cool core nor morphologically disturbed. }}\label{fig:KK500}
\end{figure*}

\subsection{Scaling}

In order to compare cluster profiles on a common radial scale, we express them in terms of $R_\delta$, the radius within which the mean mass density is $\delta$ times the critical density at the cluster redshift\footnote{$M_{\delta} = \delta \rho_c (z)\, (4\pi/3) R_{\delta}^3$, where $\rho_{\rm c}(z)= E^{2}(z)\, 3 H_0^2 / 8 \pi G$.}. For practical purposes, we generally scale to $R_{500}$, the effective limiting radius for high quality observations from \xmm\ and {\it Chandra}. Since the sample contains systems in a variety of dynamical states, we use $Y_X$ as a mass proxy. We estimate $\Rv$ iteratively as described in \citet{kvn06}, from the updated calibration of the $M_{500}$--$Y_X$ relation obtained by combining the \citet{app07} results from nearby relaxed clusters with \rexcess\ data from morphologically relaxed systems. The full sample of 20 objects (8 from \citealt{app07} and 12 from \rexcess) is comprised of all systems for which the mass profiles are measured at least down to a density contrast $\delta=550$. The resulting $M_{500}-Y_X$ relation is:

{\small
\begin{equation}
E(z)^{2/5}\Mv = 10^{14.567 \pm 0.010} \left[\frac{\YX}{2\times10^{14}\,{\msol}\,\keV}\right]^{0.561 \pm 0.018}\,{\rm h_{70}^{-1}\, \msol},\label{eqn:Yx}
\end{equation}
}

\noindent consistent with that derived by \citet{app07} but with improved accuracy on slope and normalisation \citep{arnaud09}. We also use the spectroscopic temperature $T$, measured in the $[0.15-0.75]\,\Rv$ region, to investigate the scaling properties of the entropy and associated profiles. These values are given in Table~\ref{tab:Kscaled}.

The right hand panel of Figure~\ref{fig:Kprof} shows the entropy profiles plotted in terms of $\Rv$. Plotting them this way explicitly shows the temperature dependence of the entropy distribution in the outer regions, and that despite the wide range of central entropy values, clusters clearly become more structurally similar with increasing radius. Beyond $0.2\,\Rv$, the relative dispersion in scaled entropy profiles is constant at approximately 30 per cent.

In the pure gravitational collapse scenario, the scaled profiles of any physical quantity should coincide, and so measures of these quantities at any scaled radius should correlate with global cluster parameters such as temperature or mass. The standard self-similar scalings are $K \propto E(z)^{-4/3}\,T$ for temperature and $K \propto E(z)^{-2/3}\,M^{2/3}$ for mass. Departures from these expectations are the direct result of the action of non-gravitational processes on the thermodynamics of the ICM. For comparison with previous work on entropy scaling relations, the entropy was measured for all clusters at radii equivalent to $0.1\,R_{200}$, $R_{2500}\ (\approx 0.45\,R_{500})$ and $R_{1000}\ (\approx 0.72\,R_{500})$, estimated from the scaling relations presented in \citet{app05}, via interpolation in the log-log plane. In addition, if $\Rv$ falls within the radial range encompassed by the centre of the outer temperature annulus (13 systems), we also calculated the entropy at $\Rv$. Uncertainties on the entropy were estimated from the quadratic sum of the errors associated with the deconvolved density and temperature profiles. These values are listed in Table~\ref{tab:Kscaled}.

\begin{figure*}[]
\centering
\includegraphics[scale=1.,angle=0,keepaspectratio,width=0.9\textwidth]{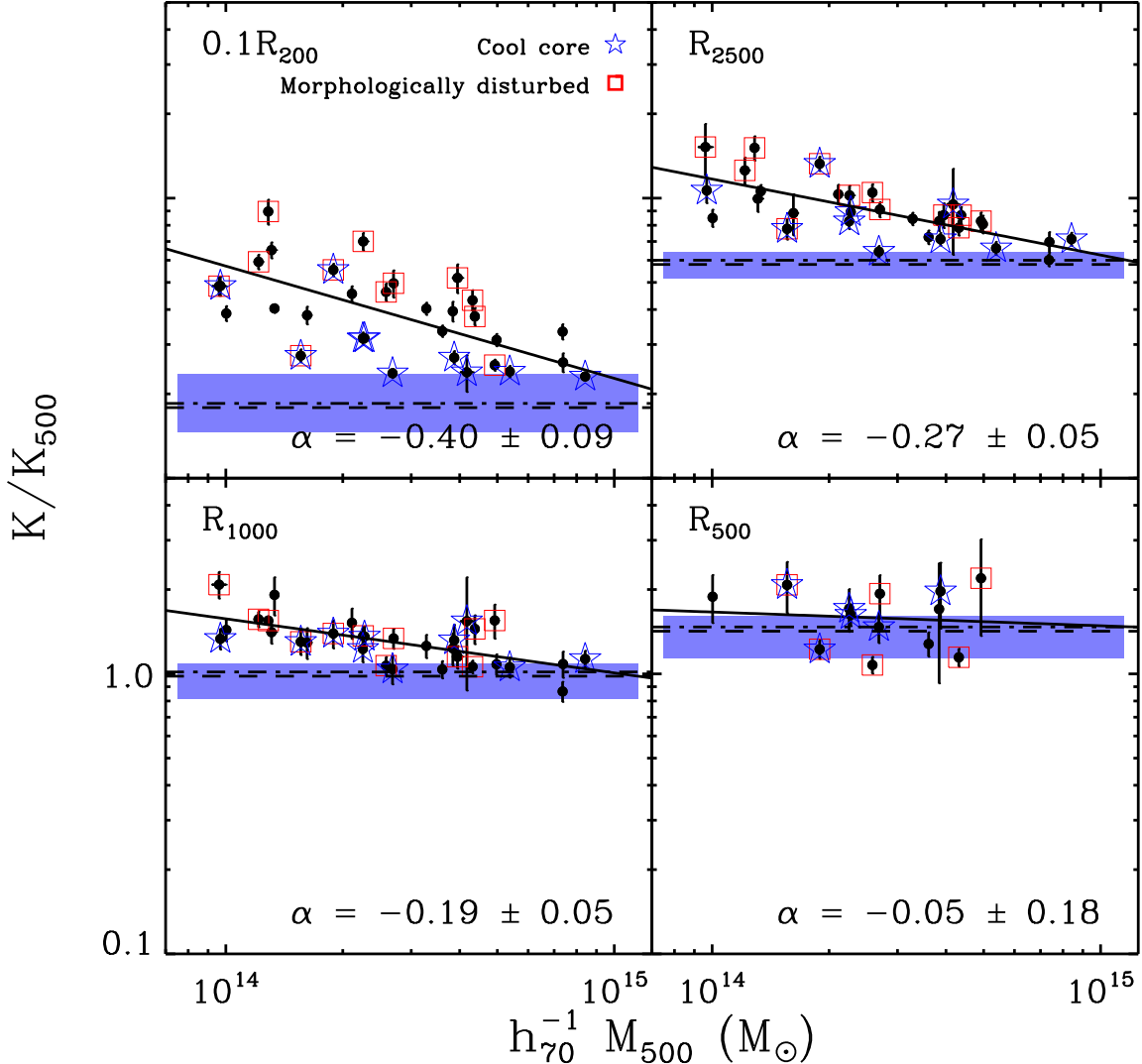}
\caption{{\footnotesize Dimensionless entropy $K/K_{500}$ as a function of mass $\Mv$ for different fractions of scaled radius. The solid lines show the power law fit to the data at each scaled radius with exponent $\alpha$; the dashed lines give the expected values from the power law relation $K/K_{500} = 1.42\, (R/\Rv)^{1.1}$ fitted to the non-radiative numerical simulations of \citet{vkb05}. The band indicates the typical dispersion of the simulated profiles about this relation at the radii indicated. The dot-dashed lines show the same relation adjusted for a 13 per cent mass underestimate due to the assumption of hydrostatic equilibrium. Cool core and morphologically disturbed subsamples are indicated.  
There is a clear segregation in central entropy properties of the subsamples at $0.1\,R_{200}$, which disappears at larger scaled radius. The dimensionless entropy at $\Rv$ is slightly higher than, but approximately consistent with expectations from simple gravitational collapse, independent of mass.}}\label{fig:MKK500}
\end{figure*}


\section{Comparison to theoretical expectations}

\subsection{Normalisation}

We first address the question of the {\it absolute} entropy normalisation with respect to theoretical expectations in the case of purely gravitational structure formation. Any deviations from the predicted normalisation would point to the influence of non-gravitational processes on the thermodynamics of the ICM. Furthermore, radially resolved entropy profiles can be used to assess the radial extent of the change in ICM properties due to non-gravitational processes, if any. In the following, we compare the normalisation of our observed entropy profiles to theoretical expectations in order to elucidate the mass dependence and radial extent of these effects. 

Numerical simulations which only implement gravitational processes make very specific predictions for both the normalisation and shape of galaxy cluster entropy profiles. \cite{vkb05} discuss such simulations, pointing out that once scaled by the characteristic entropy of the halo, 

\begin{equation}
K_{200} = \frac{1}{2} \left[ \frac{2 \pi}{15} \frac{G^2 M_{200}}{f_b H(z)} \right]^{2/3},
\end{equation}

\noindent where $f_b$ is the baryon fraction, the simulated SPH profiles, when fitted in the $[0.1-1]\,R_{200}$ radial range, scatter about a median scaled profile described by the baseline power law relation $K(R)/K_{200} = 1.32\, (R/R_{200})^{1.1}$ with approximately 20 per cent dispersion. We note that non-radiative simulations do not predict power law profiles down to arbitrarily small radii; in fact, significant flattening is generally found interior to $R \lesssim 0.2\,\Rv$. In addition, grid-based simulations consistently find larger entropy cores than SPH simulations even in the non-radiative case, which appears to be due to differences in particle mixing between the two different computational methods \citep{mitchell09}. However, beyond $\sim 0.1\,R_{200}$, the agreement between simulations is better than 10 per cent \citep{vkb05,mitchell09}. In the following, since we only compare at and beyond $0.1\,R_{200}$, we compare to the SPH results only.

Assuming an abundance of $Z=0.3\,Z_\odot$ and mean molecular weight $\mu = 0.596$, the characteristic entropy can be rewritten:

{\footnotesize
\begin{equation}
K_{500} = 106 \ {\rm keV\ cm}^{-2} \left( \frac{M_{500}}{10^{14}\,h_{70}^{-1}\,M_\odot} \right)^{2/3}\, \left(\frac{1}{f_b}\right)^{2/3}\, E(z)^{-2/3}\, h_{70}^{-4/3}.\label{eqn:K500}
\end{equation}
}

\noindent For our assumed $f_b = 0.15$ and $R_{500}/R_{200} = 0.659$ derived from an NFW profile with $c_{500}=3.2$ \citep[the mean $c$ measured for a morphologically relaxed cluster sample by][]{pap05}, the baseline relation becomes 

\begin{equation}
K(R)/K_{500} = 1.42\,(R/\Rv)^{1.1}.\label{eqn:KR}
\end{equation}

\noindent Note that this baseline relation was derived from simulations where the true masses were known, while our masses are calculated from an $M$--$Y_X$ relation derived from hydrostatic mass estimates of morphologically relaxed systems. \citet[][and references therein]{piff08} and \citet{arnaud09}, applying hydrostatic mass estimates to a large number of simulated clusters, argue that such masses can be underestimated by $-13\pm16$ per cent on average. In this case the normalisation factor in Eqn.~\ref{eqn:KR} becomes 1.47. 

\begin{table*}[]
\begin{center}
\caption{{\footnotesize Best fitting parameters for the entropy-temperature and entropy-mass relations. \label{tab:KT}}}
\centering
\begin{tabular}{l r l l l}
\hline
\multicolumn{1}{l}{Radius} & \multicolumn{1}{l}{$C$} & \multicolumn{1}{l}{$\alpha$} & \multicolumn{2}{c}{$\sigma_{\ln{K}}$} \\
\cline{4-5}
\multicolumn{1}{l}{ } & \multicolumn{1}{l}{(keV cm$^{-2}$)} & \multicolumn{1}{l}{} & \multicolumn{1}{l}{raw} & \multicolumn{1}{l}{int} \\
\hline
\\
Entropy-temperature relation & \\
\cline{1-1}
\\
$0.1\,R_{200}$ & $347\pm23$ & $0.89\pm0.15$ & $0.262\pm0.040$ & $0.254\pm0.041$ \\
$R_{2500}$ & $783\pm15$ & $0.76\pm0.06$ & $0.120\pm0.021$ & $0.083\pm0.118$ \\
$R_{1000}$ & $1152\pm27$ & $0.83\pm0.06$ & $0.093\pm0.015$ & \ldots \\
$\Rv$          & $1489\pm125$ & $0.92\pm0.24$ & $0.265\pm0.055$ & \ldots \\ 
\\
Entropy-mass relation & \\
\cline{1-1}
\\
$R_{2500}$ & $864\pm27$ & $0.42\pm0.05$ & $0.155\pm0.025$ & $0.136\pm0.031$ \\
$R_{1000}$ & $1308\pm52$ & $0.48\pm0.04$ & $0.119\pm0.017$ & $0.052\pm0.162$  \\
$\Rv$      & $1748\pm237$ & $0.62\pm0.17$ & $0.265\pm0.055$ & $0.221\pm0.160$ \\ 
\\
\hline
\end{tabular}
\end{center}
{\footnotesize $T$ is the spectroscopic temperature in the $[0.15-0.75]\,\Rv$ region; masses are estimated from the $M-Y_X$ relation given in Eqn.~\ref{eqn:Yx}. Data were fitted with a power law of the form $E(z)^{n} K = C \times (A/A_0)^\alpha$, with $A_0 = 5$ keV and $5.3\times10^{14}\,M_\odot$, and $n= 4/3$ and $2/3$, for temperature and mass repectively. Fits used orthogonal BCES regression with errors estimated using bootstrap resampling.  The raw and intrinsic logarithmic scatter about the best fitting relations are given in the final two columns.}
\end{table*}

Figure~\ref{fig:KK500} shows the dimensionless entropy profiles $K(R)/K_{500}$. Clearly, the observed profiles do not coincide. The central regions show the most dispersion and the profiles tend to converge towards the non-radiative prediction at large radius, but those of the lowest mass systems converge slowest; in other words, their slopes are shallower (discussed in more detail below). This is a manifestation of the fact that the entropy modification extends to larger radii in lower temperature systems, consistent with the expectation that non-gravitational processes have a greater effect at the low mass end of the cluster population. 

To better quantify the above, Figure~\ref{fig:MKK500} shows the dimensionless entropy $K/K_{500}$ versus mass for various fractions of scaled radius. Also overplotted in each panel is the expectation from Equation~\ref{eqn:KR}. As expected, it can be seen that at $0.1\,\Rv$, the excess with respect to the theoretical prediction from gravitational collapse is strongly mass-dependent, with the least massive systems exhibiting the strongest deviation. 
The mass dependence becomes less pronounced as we proceed towards the outer regions of the ICM, such that at $\Rv$, the mass dependence is entirely consistent with zero and the measured values scatter about the theoretical prediction. Indeed, at this radius, the very slight negative slope can be attributed to the single lowest mass data point, which drives the fit.

At small radii there is large amount of scatter ($\sigma_{\ln{K},{\rm int}} = 0.26\pm0.04$ about the best fitting regression line). Dividing the data into subsamples elucidates the origin of this scatter: there is a clear segregation in subsamples, with cool core systems showing the least deviation from the baseline prediction while morphologically disturbed systems show the most deviation. The subsample segregation disappears as we push outward though, and there is no evidence for any segregation at or beyond $R_{1000}$. The full radial behaviour of this trend is explicitly illustrated in the right hand panel of Figure~\ref{fig:KK500}.

Interior to $\Rv$, the observed entropy is always higher than the baseline prediction. However, at $\Rv$, the median dimensionless entropy is $K(\Rv)/K_{500} = 1.70 \pm 0.35$, where the uncertainty comes from the standard deviation of the points. This is slightly higher than, but consistent with both the baseline prediction of $K(\Rv)/K_{500} = 1.42$ (Eqn.~\ref{eqn:KR}) and the same prediction corrected for a 13 per cent mass bias due to the assumption of hydrostatic equilibrium, $K(\Rv)/K_{500} = 1.47$. The lack of mass dependence and agreement with the normalisation from simulations was also noted for a sample of cool core clusters by \citet{nkv07}. For the present representative sample, given the large uncertainties it is not possible to test the predictions more thoroughly, underlining the need for robust, high quality, spatially resolved entropy measurements at and beyond $\Rv$. 


\subsection{Entropy scaling relations}

For comparison with previous work, we also examined the entropy-temperature and entropy-mass relations. Scaling relations were fitted with a power law of the form $E(z)^n B = C(A/A_0)^\alpha$, with $A_0= 5$ keV and $5.3 \times 10^{14}\,M_\odot$ for $T$ and $M$ respectively, and $n$ fixed to the expected scaling with redshift ($n=4/3$ for $T$ and $2/3$ for $M$). Data were fitted using the orthogonal BCES minimisation technique \citep{bces} with uncertainties on each fit parameter estimated from bootstrap resampling.

The best fitting slopes and intercepts for the entropy-temperature and entropy-mass relations at various scaled radii are listed in Table~\ref{tab:KT}. The evolution of these slopes with increasing radius mirrors the behaviour of the dimensionless entropy discussed above; in the inner regions the relations are shallower than self-similar with large scatter, while at $\Rv$ the relations are compatible with self-similar (although with relatively large uncertainties given the limited number of data points). 

Comparing to previous work, a wide variety of slopes have been found from fits to the entropy-temperature relation at $0.1\,R_{200}$, ranging from very shallow ($\alpha = 0.49\pm0.15$: \citealt{pap06}; $\alpha= 0.50\pm0.08$: \citealt{nkv07}) to very steep ($\alpha=0.92\pm0.12$: \citealt{sanderson09}; $\alpha=0.85\pm0.19$: this work). We simply note that cool core-only samples tend to yield shallower slopes than statistically-selected samples, a fact borne out in the present data, for which the entropy temperature relation at $0.1\,R_{200}$ for the cool core subsample has a slope of $\alpha=0.63\pm0.94$, while the morphologically disturbed subsample has a slope of $\alpha=1.22\pm0.76$. Beyond $0.1\,R_{200}$, both the slope and the normalisation of the relations are in very good agreement with recent determinations \citep[][]{nkv07,sun09}, showing the excellent consistency between \xmm\ and {\it Chandra} results\footnote{Our earlier results suggested somewhat shallower relations at $R_{1000}$ \citep{pap06}. The difference can be traced to the increased precision on the temperature profiles afforded by the present data, especially at low mass.}.

None of the studies listed above give constraints on entropy evolution. In this context, our results underline the need for representative samples, to establish the effects of non gravitational processes and dynamical state on the evolution in the central regions. Furthermore, precise measurements at large radius ($R \gtrsim R_{2500}$) are needed to establish the baseline entropy evolution in the absence of non-gravitational effects.

\begin{figure*}[]
\begin{centering}
\includegraphics[scale=1.,angle=0,keepaspectratio,width=0.45\textwidth]{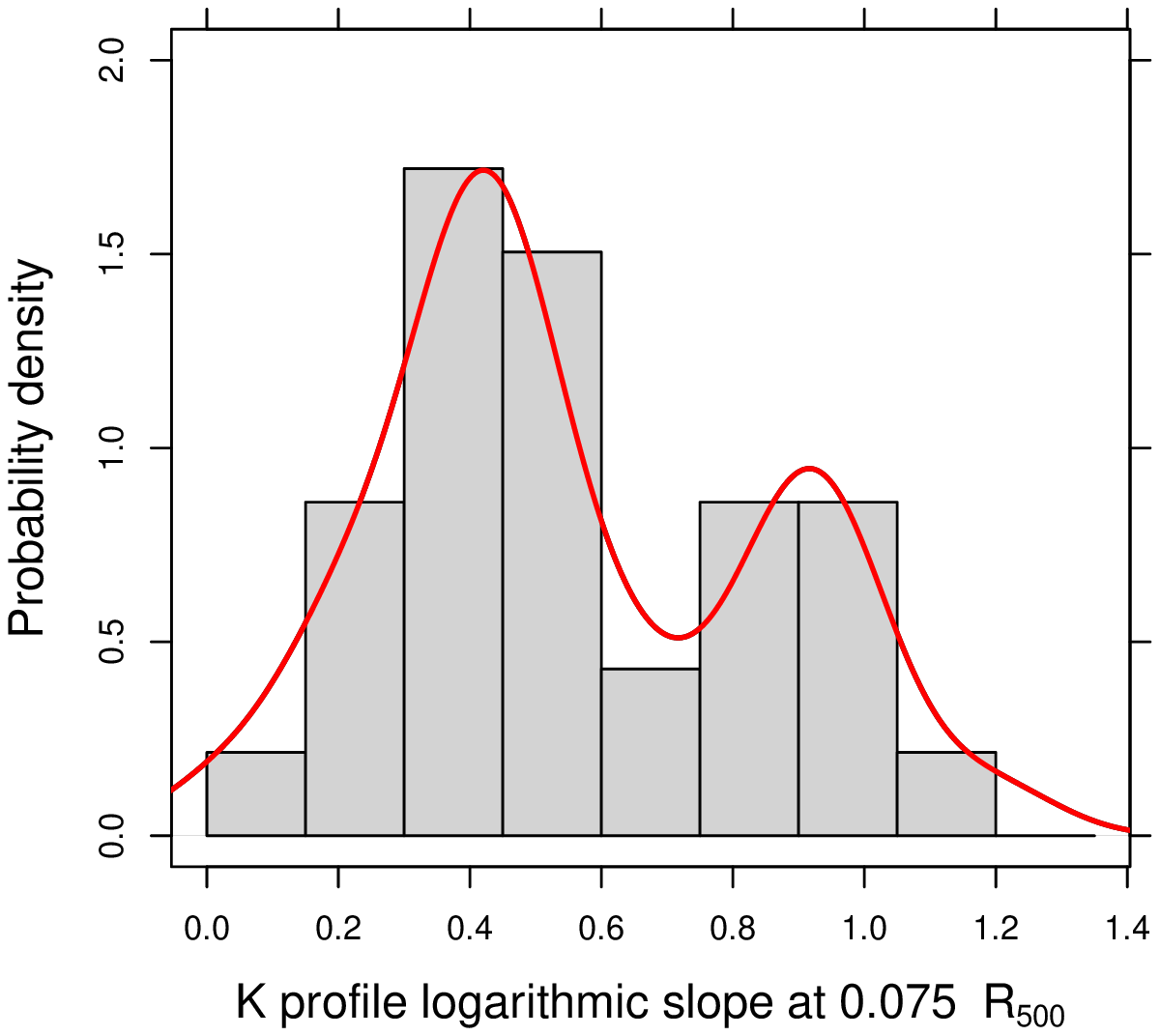}
\hfill
\includegraphics[scale=1.,angle=0,keepaspectratio,width=0.45\textwidth]{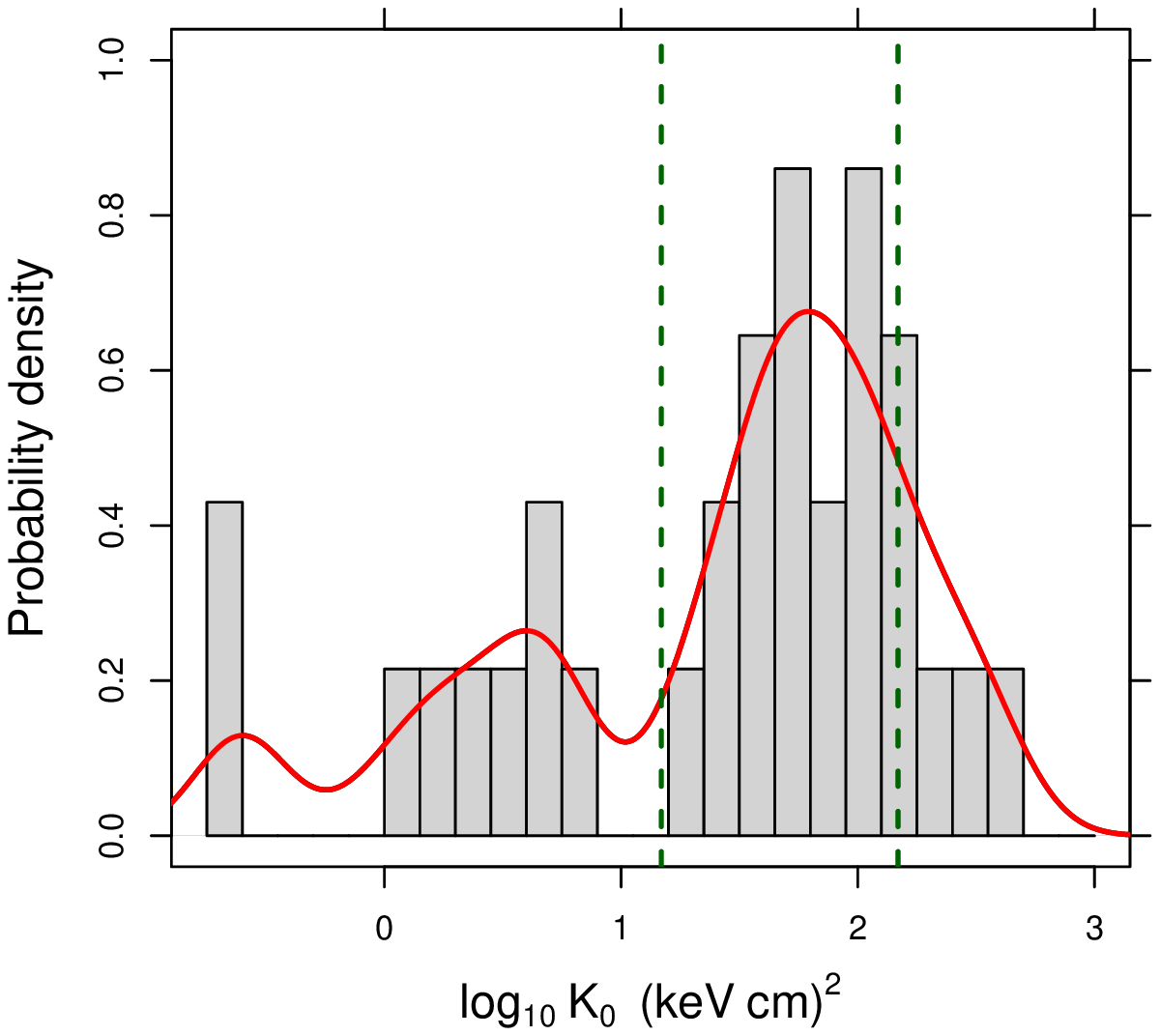}
\end{centering}
\caption{{\footnotesize Central regions of the \rexcess\ entropy profiles. {\it Left panel:} Probability density plot of the logarithmic slope of the density profile measured at a fiducial radius of $0.075\,\Rv$. Clusters with a steep slope correspond to cool core systems. The solid line is a kernel density plot with a smoothing width of 0.15. {\it Right panel:} Probability density plot of the central entropy excess above a power law, $K_0$, from a power law plus a constant model fit $K(R) = K_0 + K_{100}\, (R/100\, h_{70}^{-1}\, {\rm kpc})^\alpha$, to the entropy profiles. The solid line is a kernel density plot with a smoothing width of 0.1. Cool core clusters have the smallest values of $K_0$.}}\label{fig:khist1}
\end{figure*}


\section{Radial entropy structure}

Various semi-analytical models and cosmological simulations of clusters formed in the absence of non-gravitational processes have shown that outside the central regions ($R > 0.1\,R_{200}$, or $\sim 0.15\,\Rv$) , entropy profiles follow a power law with $K(R) \propto R^{1.1}$ \citep{tn01,borg05,voit02,vkb05,mitchell09}. Simulated profiles flatten in the very central regions due to entropy mixing \citep{wad08,mitchell09}. Observed profiles are also found generally to have similar external slopes \citep[e.g.,][]{pap06,sun09} and to flatten in the central regions in high resolution {\it Chandra} observations \citep{donahue06,sanderson09,cav09}. In the following, we investigate the central entropy slope and a parameterisation of the entropy profiles, and relate this to global cluster properties.


\subsection{Central slope}
\label{sec:slope}

The left hand panel of Figure~\ref{fig:khist1} shows the probability density distribution of the logarithmic slope $d \ln{K(R)}/d \ln{R}$ of the entropy profiles measured at $0.075\,\Rv$, where we have directly measured data for all but two systems. The distribution shows two peaks, with the larger peak at a slope of $\sim 0.4$, containing approximately 2/3 of the sample,  and a smaller peak at a slope of $\sim 0.9$, which comprises the cool core systems defined in the subsample classification scheme discussed in Section~\ref{sec:sample}. Two peaks in the distribution of inner logarithmic slope were also found in the {\it Chandra} analysis of \citet[][in their case, at $0.05\,\Rv$]{sanderson09}.

\begin{figure*}[]
\begin{centering}
\includegraphics[scale=1.,angle=0,keepaspectratio,width=0.45\textwidth]{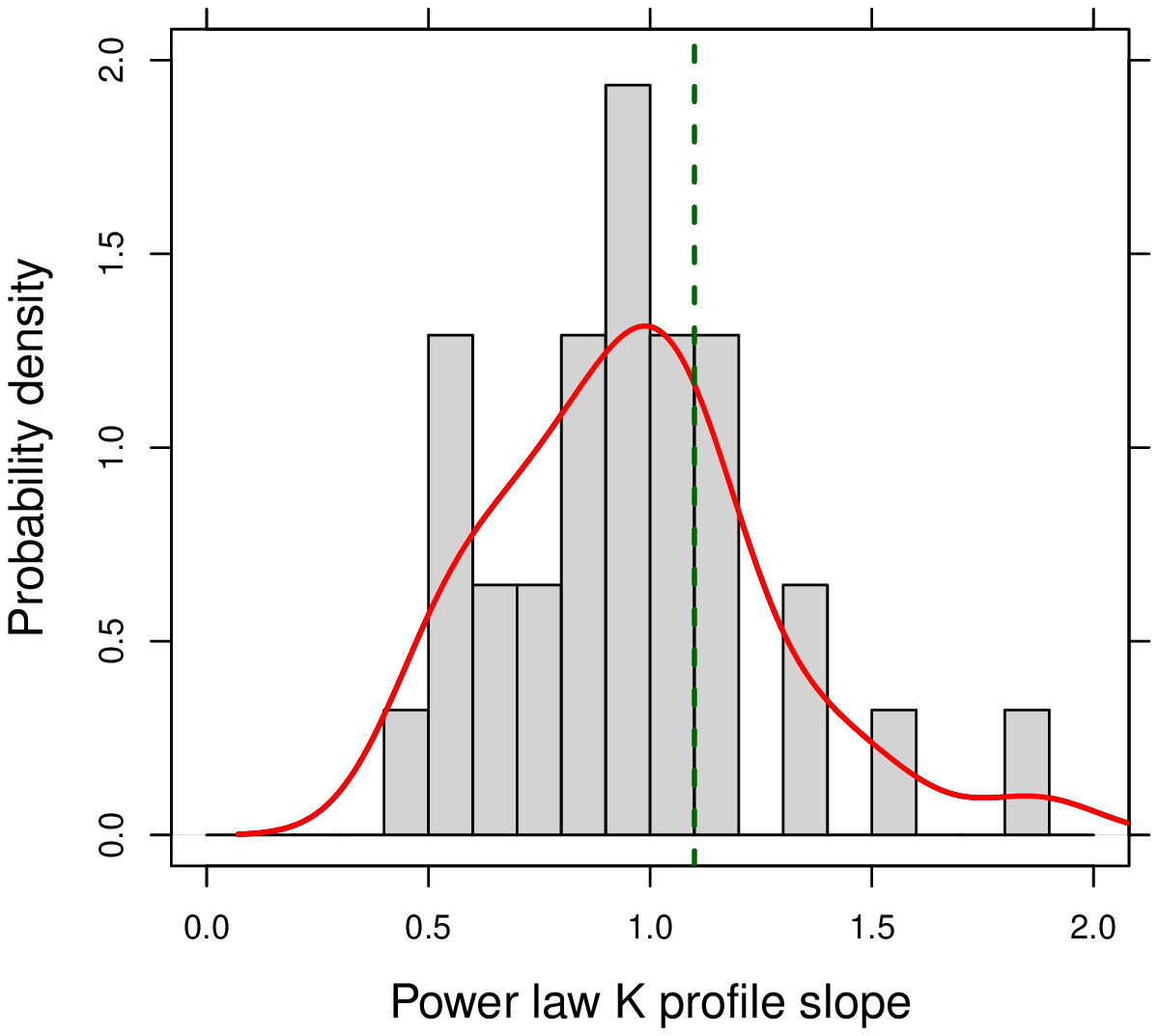}
\hfill
\includegraphics[scale=1.,angle=0,keepaspectratio,width=0.45\textwidth]{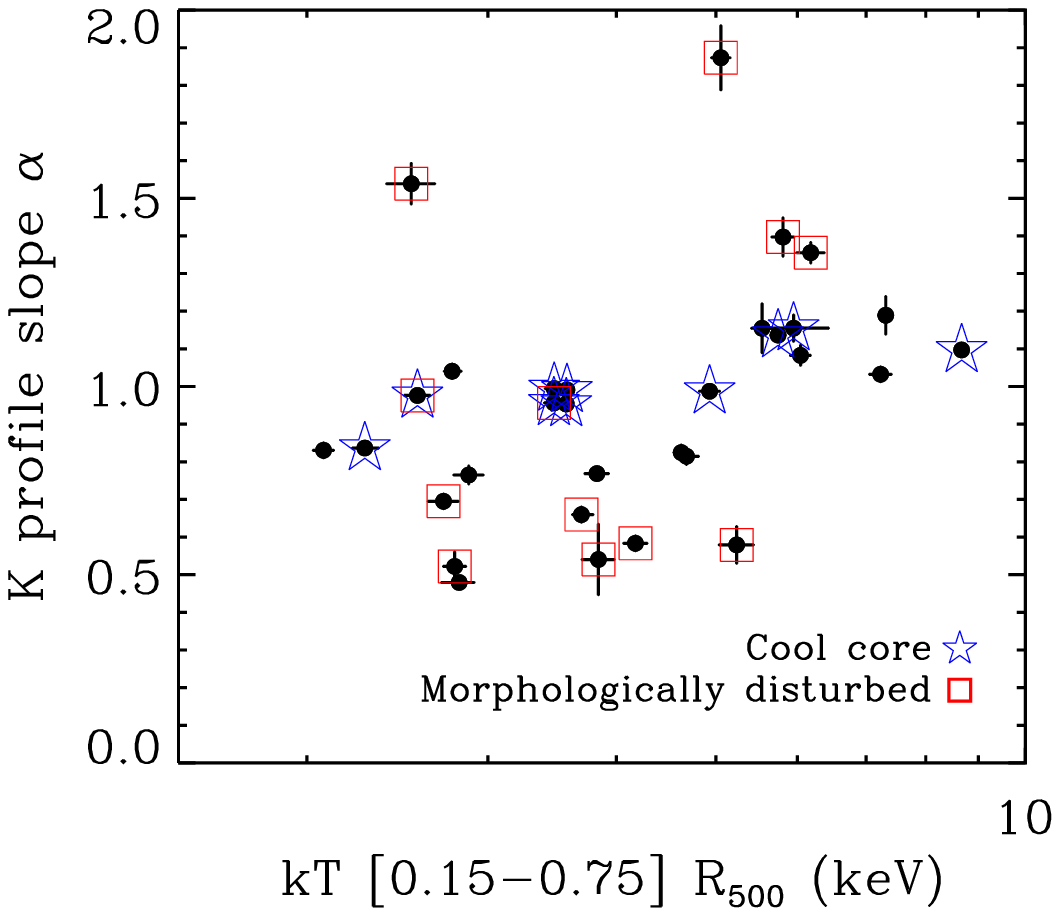}
\end{centering}
\caption{{\footnotesize Outer regions of the \rexcess\ entropy profiles. {\it Left panel:}  Probability density plot of outer slope values, $\alpha$, from a power law plus a constant model fit $K(R) = K_0 + K_{100}\, (R/100\, h_{70}^{-1}\, {\rm kpc})^\alpha$, to the entropy profiles. The dashed line shows a value of 1.1. The solid line is a kernel density plot with a smoothing width of 0.1.  {\it Right panel:} Entropy profile slope vs temperature in the $[0.15-0.75]\,\Rv$ aperture. Cool core and morphologically disturbed subsamples are indicated.}}\label{fig:khist2}
\end{figure*}


\subsection{Parameterised fitting}

\subsubsection{Model}

A power law is the simplest parameterisation of the entropy profiles, but Figure~\ref{fig:Kprof} shows that such a description is clearly inadequate to describe the majority of the profiles in the present sample if the core regions are included in the fit. A possibly more interesting parameterisation is to instead model the entropy profiles with a power law plus a constant, $K(R) = K_0 + K_{100}\, (R/100\, h_{70}^{-1}\, {\rm kpc})^\alpha$ \citep[as introduced by][]{donahue05}, where $K_0$ represents the typical excess of core entropy above the best fitting power law at large radii. In this case, to ensure that the same radial range is fitted in all cases, and to ensure the precision of the powerlaw fit at large radii, the profiles were fitted between $[R_{\rm in} - R_{1000}]$. The results of fitting the power law plus constant model are listed in Table~\ref{tab:kpar}. The external power law is extremely stable to fitting ranges and clearly does not depend on the entropy of the central region or the value of $K_{100}$. 

\begin{table}[]
\begin{center}
\caption{{\footnotesize Entropy profile parameterisation with a model of the form $K(R) = K_0 + K_{100}\, (R/100\, h_{70}^{-1}\, {\rm kpc})^\alpha$. Entropy is measured in units of keV cm$^{-2}$. \label{tab:kpar}}}
\centering
{\tiny 
\begin{tabular}{l r r r}
\hline
\hline
\\
\multicolumn{1}{l}{Cluster} & \multicolumn{1}{c}{$K_0$} & \multicolumn{1}{c}{$K_{100}$} & \multicolumn{1}{c}{$\alpha$} \\
\hline
\\
RXC\,J0003.8+0203 & $    39.33 \pm   2.20 $ & $   196.25 \pm   3.93 $ & $     0.77 \pm   0.02 $ \\
RXC\,J0006.6-3443 & $    97.68 \pm  26.02 $ & $   280.10 \pm  30.12 $ & $     0.58 \pm   0.05 $ \\
RXC\,J0020.7-2542 & $   177.45 \pm  10.39 $ & $    87.07 \pm   9.76 $ & $     1.15 \pm   0.07 $ \\
RXC\,J0049.4-2931 & $    35.95 \pm   2.32 $ & $   141.70 \pm   3.46 $ & $     0.76 \pm   0.02 $ \\
RXC\,J0145.0-5300 & $   240.05 \pm   5.49 $ & $    63.39 \pm   5.40 $ & $     1.40 \pm   0.05 $ \\
RXC\,J0211.4-4017 & $    19.38 \pm   1.58 $ & $   118.87 \pm   2.57 $ & $     0.83 \pm   0.02 $ \\
RXC\,J0225.1-2928 & $    94.57 \pm   1.93 $ & $    79.07 \pm   3.74 $ & $     1.54 \pm   0.06 $ \\
RXC\,J0345.7-4112 & $     1.03 \pm   0.44 $ & $   153.59 \pm   1.68 $ & $     0.84 \pm   0.01 $ \\
RXC\,J0547.6-3152 & $   141.30 \pm   3.21 $ & $   111.99 \pm   4.49 $ & $     1.08 \pm   0.03 $ \\
RXC\,J0605.8-3518 & $     4.78 \pm   0.67 $ & $   139.62 \pm   1.65 $ & $     0.99 \pm   0.01 $ \\
RXC\,J0616.8-4748 & $    22.46 \pm   6.00 $ & $   268.72 \pm   9.09 $ & $     0.58 \pm   0.02 $ \\
RXC\,J0645.4-5413 & $    49.35 \pm   2.65 $ & $   145.40 \pm   4.45 $ & $     1.03 \pm   0.02 $ \\
RXC\,J0821.8+0112 &       \ldots            & $   263.95 \pm   2.66 $ & $     0.48 \pm   0.01 $ \\
RXC\,J0958.3-1103 & $    24.78 \pm   1.89 $ & $   117.86 \pm   3.58 $ & $     1.15 \pm   0.04 $ \\
RXC\,J1044.5-0704 &       \dots             & $   117.65 \pm   0.53 $ & $     0.95 \pm   0.00 $ \\
RXC\,J1141.4-1216 & $     2.06 \pm   0.17 $ & $   148.75 \pm   0.76 $ & $     0.99 \pm   0.01 $ \\
RXC\,J1236.7-3354 & $    52.14 \pm   1.32 $ & $   112.12 \pm   2.18 $ & $     1.04 \pm   0.02 $ \\
RXC\,J1302.8-0230 & $     3.60 \pm   0.42 $ & $   209.91 \pm   1.92 $ & $     0.96 \pm   0.01 $ \\
RXC\,J1311.4-0120 & $    49.20 \pm   1.45 $ & $   134.89 \pm   2.10 $ & $     1.10 \pm   0.01 $ \\
RXC\,J1516+0005 & $    88.33 \pm   3.68 $ & $   165.48 \pm   5.36 $ & $     0.81 \pm   0.02 $ \\
RXC\,J1516.5-0056 & $    43.01 \pm   5.19 $ & $   229.09 \pm   7.74 $ & $     0.66 \pm   0.02 $ \\
RXC\,J2014.8-2430 & $     1.75 \pm   0.26 $ & $   121.61 \pm   1.00 $ & $     1.14 \pm   0.01 $ \\
RXC\,J2023.0-2056 & $    52.27 \pm   3.02 $ & $   194.29 \pm   4.41 $ & $     0.69 \pm   0.02 $ \\
RXC\,J2048.1-1750 & $   370.96 \pm   6.31 $ & $    14.53 \pm   2.44 $ & $     1.87 \pm   0.09 $ \\
RXC\,J2129.8-5048 & $   150.49 \pm  56.32 $ & $   232.96 \pm  56.20 $ & $     0.54 \pm   0.09 $ \\
RXC\,J2149.1-3041 & $     4.26 \pm   0.69 $ & $   136.22 \pm   1.53 $ & $     0.99 \pm   0.01 $ \\
RXC\,J2157.4-0747 & $    92.24 \pm  15.42 $ & $   294.40 \pm  19.04 $ & $     0.52 \pm   0.04 $ \\
RXC\,J2217.7-3543 & $    63.35 \pm   3.25 $ & $   155.35 \pm   4.88 $ & $     0.82 \pm   0.02 $ \\
RXC\,J2218.6-3853 & $    98.10 \pm   1.92 $ & $    80.84 \pm   2.91 $ & $     1.36 \pm   0.03 $ \\
RXC\,J2234.5-3744 & $   308.07 \pm   6.98 $ & $    62.01 \pm   5.92 $ & $     1.19 \pm   0.05 $ \\
RXC\,J2319.6-7313 & $     6.18 \pm   0.26 $ & $   105.24 \pm   0.93 $ & $     0.98 \pm   0.01 $ \\

\\
\hline
\end{tabular}
}
\end{center}
\end{table}



\subsubsection{Distribution of central entropy, $K_0$}

The histogram of $K_0$ values for the \rexcess\ sample is shown as a probability density in the right hand panel of Figure~\ref{fig:khist1}, where the bins are 0.15 dex and clusters with a $K_0$ consistent with zero (i.e., pure power law profiles) are shown at the extreme left of the plot. For the \rexcess\ sample, the number of systems with a $K_0$ consistent with zero at $3\sigma$ significance is three for a constant central temperature (10 per cent of the sample), consistent with \citet{cav09}, who analysed a large number of {\it Chandra} archive observations and who also used a constant central temperature assumption.

\citet[][]{cav09} found that the distribution of central entropies in their sample is bimodal with peaks of approximately the same amplitude at $K_0 \sim 15$ keV cm$^2$ and $K_0 \sim 150$ keV cm$^2$, and a distinct gap between $K_0 \sim 30 - 50$ keV cm$^2$ \citep[see also][]{hr07}. The peak values are indicated in Figure~\ref{fig:khist1} by a dashed line. The distribution of \rexcess\ central entropies also exhibits two peaks, although there are subtle differences in their positions and amplitudes. In particular, there are more clusters with a high $K_0$ than with a low $K_0$, and the positions of both peaks are shifted somewhat to lower $K_0$ with respect to those found by \citeauthor{cav09}


\subsubsection{Entropy slope outside the core}

The left hand panel of Figure~\ref{fig:khist2} shows the histogram of fitted values of the outer profile slope $\alpha$ obtained from the power law plus constant model. While there is a quite substantial spread of values in entropy profile slope, ranging from extremely shallow ($\alpha \sim 0.5$) to extremely steep ($\alpha \sim 1.9$), there is no indication for bimodality in the distribution of outer entropy slope when the profiles are modelled in this way. The median slope for the power law plus constant model fits is $0.98$, which is slightly shallower than the canonical value of 1.1 (indicated by the dashed line in the Figure). It is lower still than the value of 1.2 which was found by \citet{vkb05} when fitting the $[0.2-1]\,R_{200}$ radial range.

The right hand panel of Figure~\ref{fig:khist2} shows the distribution of slopes versus temperature. The Spearman rank correlation coefficient is 0.53 with a significance of $2.2 \times 10^{-3}$ indicating a significant correlation of slope with temperature. This is most likely a manifestation of the well-known dependence of outer density profile slope on temperature \citep[see][for the present sample]{croston08}. The different subsamples are indicated in the Figure by blue stars (cool core systems) and red squares (morphologically disturbed systems). Clearly the cool core subsample has a very small scatter in outer entropy slope; in addition there is a more pronounced trend with temperature, illustrated by the fact that the Spearman rank coefficient for this subsample is  0.77 with a significance of $9.7 \times10^{-3}$. The morphologically disturbed subsample has a very large scatter, incorporating both the upper and lower extremes in outer slope values. The tight range of slopes in the cool core systems, together with the wide range of slopes in the morphologically disturbed systems,  suggests that the core and external properties of the ICM are linked.


\section{Discussion}

\subsection{Bimodality?}

Two peaks are seen in the central regions of the \rexcess\ sample, which are visible both in the distribution of logarithmic entropy slope at $0.075\Rv$ and in the distribution of $K_0$ from parameterisation of the profiles with a power law plus constant model (Figure~\ref{fig:khist1}). For a more quantitative comparison with the results of \citet{cav09}, we performed a maximum likelihood fit of the 29 clusters for which $K_0$ is constrained. The fitting was performed on the unbinned data in log space, using the {\sc mclust} and {\sc fitdistr} packages in version 2.9 of the {\sc r} statistical software environment\footnote{{\tt http://www.r-project.org}}. We considered three different models: a single Gaussian, a left-skewed distribution, and a double Gaussian (i.e., a bimodal distribution), and used the Bayesian information criterion \citep[BIC,][]{sch78} to distinguish between them. A difference in BIC of between 2 and 6 indicates positive evidence against the model with the greater BIC value; values above 6 indicate strong evidence against the model with the greater BIC value. The BIC values are -67.53, -63.32 and -63.08 for the single, left-skewed and double Gaussian distributions, respectively. Thus while the single Gaussian distribution is clearly the worst description of the data, on the basis of this test, and given the limited number of data points at our disposal, we cannot definitively distinguish between a bimodal and a left-skewed distribution of $K_0$. 

The best bimodal distribution fit yields Gaussian means at $K_0 \sim 3$ and $K_0 \sim 75$ keV cm$^2$, with an amplitude ratio of 1:3. If we associate these values with the peaks found by \citet{cav09}, then they are somewhat offset to lower values and the amplitude ratio is different (\citeauthor{cav09} found peaks at $\sim 15$ and $\sim 150$ keV cm$^2$, with an amplitude ratio close to 1:1). The shift of the higher central entropy peak is due to a number of clusters in the \rexcess\ sample that fall directly in the $K_0 \sim 30-50$ keV cm$^2$ gap found by \citeauthor{cav09} It may be that these clusters have a more `typical' value of $K_0$ for the general population. Given that the \citet{cav09} sample was archive limited, and given that prevailing sociological trends in cluster research have for many years led to a focus on extreme cool cores and spectacular mergers to the exclusion of more mundane objects, it may well be that clusters with a more `typical' central entropy do not exist there but do exist in a representative sample. We note however that \citeauthor{cav09} also detected bimodality in a complete flux-limited subsample of their data (although at much reduced significance), with the peaks shifted slightly lower relative to those of the full archive sample. Alternatively, the recent 2009 January {\it Chandra} calibration update may offer a more prosaic, albeit partial, explanation. Since the \citeauthor{cav09} analysis predates this update, hotter systems with flat central temperature distributions ($kT \gtrsim 4$ keV), would have a systematically higher temperature, and hence entropy.

The shift of the lower central entropy peak appears to be due to technical differences connected to the treatment of temperature profiles. \citeauthor{cav09} derived their entropy profiles from {\it Chandra} data, which has sub-arcsecond resolution in the centre of the detector, meaning that no PSF correction was needed. However, while their density profiles were deprojected, their temperature profiles were not. For flat central temperature distributions, this will not substantially change the resulting entropy. However, for steeply declining central temperature distributions, neglect of projection effects will tend to lead to an overestimate of the temperature of the inner annulus  \citep[see e.g., Fig.~8 of][]{pointeco04}, and thus of the central entropy for the assumption of constant central temperature. The net effect of neglecting deprojection would be to shift the peak in $K_0$ to higher values in the {\it Chandra} analysis, as observed.

We note that the differences in the central entropy distributions of \rexcess\ and \citet{cav09} are unlikely to be due to resolution effects. We extracted surface brightness profiles in 3\farcs3 bins while \citet{cav09} used 5\arcsec\ bins, meaning that for a given redshift, our \xmm\ entropy profiles in fact have a higher resolution than those from {\it Chandra}. Our 3\farcs3 bins give us a resolution of 3.5-10 $h_{70}^{-1}$ kpc for the present sample, depending on redshift, with a median value of 7 $h_{70}^{-1}$ kpc. Given that entropy cores extend typically to 30 kpc in {\it Chandra} observations of cool core systems \citep{donahue05}, these should be easily detectable with the current data if they exist. However, {\it Chandra} follow-up observations would still be desirable to quantify the cores of these systems at higher resolution, and particularly to investigate if there is indeed a turnover in entropy at very small radius in systems where we are unable to detect this effect.

We also do not find any bimodality in the distribution of outer power law slope $\alpha$, consistent with the increased self-similarity of the profiles in the outer regions. In this context it is important to underline the fact that our fitted model consisted of a power law plus an additive constant, and is thus different from the simple power law model fitted by e.g., \citet[][]{sanderson09}.


\subsection{Maintaining the distribution of central entropies}

The clear link between the lack of a cool core and the presence of morphological disturbance established by the \rexcess\ sample gives important clues to the processes at play in giving rise to the observed distribution of central entropies. It seems paradoxical that the profiles of cool core systems, where non-gravitational processes play an important role, resemble most closely the non-radiative baseline, while those of the unrelaxed objects, whose properties are expected to be dominated by gravitational processes, deviate the most from the baseline.

A possible explanation for the form of the cool core profiles is that in these systems, AGN heating is gentle and serves primarily to balance cooling, thus preserving the increasing form of the entropy profiles. If this were indeed the case, it would imply that AGN heating is achieved via mechanisms such as weak shocks \citep[e.g.,][]{fabian03}, or buoyant bubbles \citep[e.g.,][]{churazov01}, rather than via catastrophic explosions.

We can envisage two scenarios that could explain the properties of the morphologically disturbed clusters in \rexcess\ and the link between the lack of a cool core and the evidence that a cluster is unrelaxed. One explanation is that a combination of extra heating and dynamical activity have conspired to keep the central entropy elevated and so prevent the morphologically disturbed systems in the \rexcess\ from ever forming a cool core in the first place. This is possible because post merger disturbance persists for longer in these clusters, as in a disrupted core with less entropy contrast, buoyancy differences will not be as strong, and the stratification that restores a cluster to a relaxed state will happen more slowly. 

The second explanation is simply that the morphologically disturbed systems in the \rexcess\ were originally cool core systems, and that mergers have disrupted the entropy structure in the central regions. Given time, they should relax back to a cool core state, but we are seeing them before the cool core can re-establish itself. We note that such a scenario is supported in terms of timescales by the observed relationship between the lack of cool cores, clear evidence of dynamical disturbance and the presence of radio halo emission \citep[e.g.][]{govoni04}. The ongoing \rexcess\ radio follow-up will be extremely useful in determining if this possibility can explain the clear anti-correlation of cool cores and disturbed morphology. 

\begin{figure}[]
\begin{centering}
\includegraphics[scale=1.,angle=0,keepaspectratio,width=1.05\columnwidth]{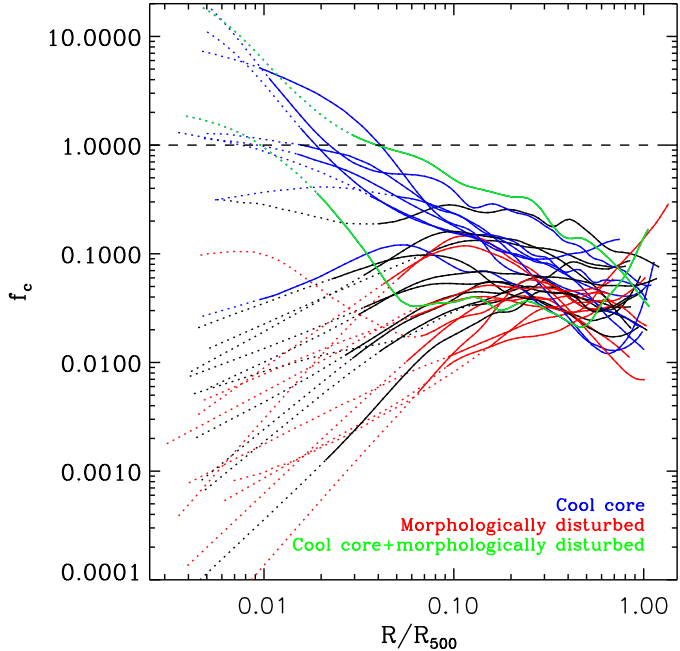}
\end{centering}
\caption{{\footnotesize Implied Spitzer conductivity suppression factor as a function of radius. The different cluster subsamples are indicated; line styles are as for Figure~\ref{fig:Kprof}. The dashed line indicates the threshold between thermally unstable and conductively stable regimes.}}\label{fig:fc}
\end{figure}
\begin{figure}[]
\begin{centering}
\includegraphics[scale=1.,angle=0,keepaspectratio,width=1.05\columnwidth]{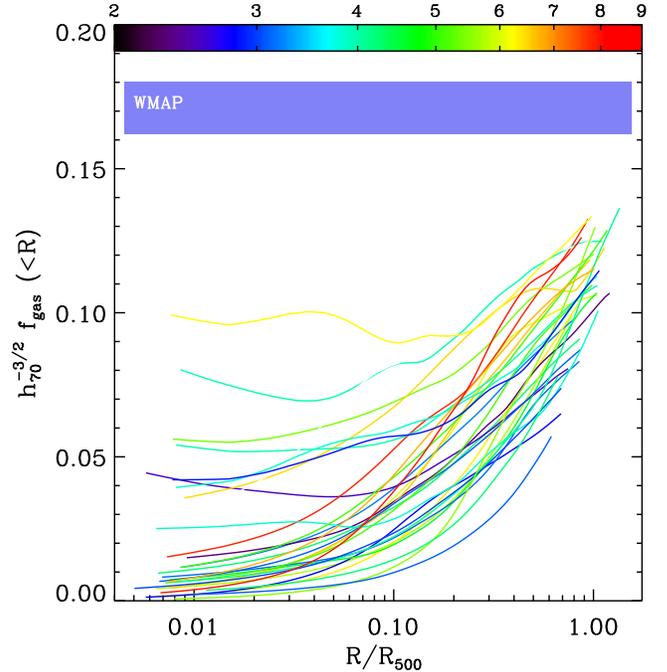}
\end{centering}
\caption{{\footnotesize Gas mass fraction profiles $f_{\rm gas} (< R)$. Total mass profiles are calculated for an NFW profile with a concentration $c_{500}=3.2$ \citep{pap05}, normalised to the $\Mv$ estimated from Eqn.~\ref{eqn:Yx}.}}\label{fig:fg}
\end{figure}

Furthermore, it is possible that the timescale for cooling to (re-) establish itself is impaired by conduction. Figure~\ref{fig:fc} shows the conduction suppression factor $f_c$ expressed in terms of the radius (in kpc) and the entropy (in keV cm$^2$), viz., $f_c \approx 62.5\, (R^2/K^3)$ \citep[][]{sanderson09}\footnote{Note that this equation only holds at the limit of conductive thermal balance, i.e. where the radius exactly matches the Field length at all radii.}.  It is clear that non-cool-core clusters can be stabilised since conductivity would need to be suppressed at most by only a factor of $f_c \sim 0.1$ to locally counteract cooling (Figure~\ref{fig:fc}), while this will not be effective in the centres of most cool core systems, where the threshold for conductive stability ($f_c = 1$) is exceeded (a similar result was found by \citealt{sanderson09}). Thus conduction could also contribute to sustaining the elevated central entropy in post-merger systems long after the initial disruption.

We note that high central entropy non-cool-core systems are difficult to reproduce in numerical simulations, which seems to be a consequence of the presence of very dense cores which are very hard to disrupt \citep[e.g.][]{gomez02,poole08}. The presence of these dense cores, of which in some cases multiple instances may be present in the same system (B\'ohringer et al., in preparation), at variance with observations, may point to deficiencies in the modelling of the complex interplay between  gravitational and non-gravitational processes in these simulations.


\subsection{Linking entropy and gas mass fraction}

Entropy modification is generally discussed in terms of three basic mechanisms: early heating (`pre-heating'), where the gas is heated before accretion into the dark matter potential well, presumably either by early supernovae and/or AGN activity \citep[e.g.][]{kaiser91,evrard91};  internal heating after accretion by the same or similar mechanisms \citep[e.g.][]{me94,bower08}; radiative cooling of the gas \citep[e.g.][]{pearce00}, where the lowest entropy gas found in the centre of the cluster condenses and cools out of the ICM. 
All of these processes act to change the total amount of gas in the central regions of a cluster, either through making it more difficult to compress into the halo (early heating), through convection of gas to the outer regions (internal heating), or through physical removal of the gas to form stars (cooling).  

\begin{figure}[]
\begin{centering}
\includegraphics[scale=1.,angle=0,keepaspectratio,width=1.05\columnwidth]{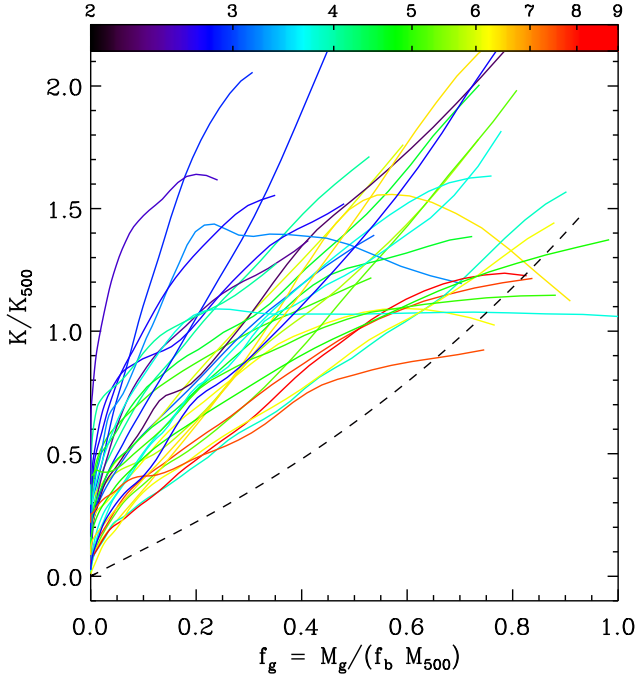}
\end{centering}
\caption{{\footnotesize The dimensionless entropy $K/K_{500}$ rises with $f_g (R) = M_g / (f_b \Mv)$, the fraction of a cluster's baryons in the ICM within radius $R$ in all clusters. However, the entropy distribution shows a clear temperature dependence. The dashed line illustrates the unmodified entropy distribution for a cluster of $\Mv = 8 \times 10^{14}\, M_\odot$ (approximately the mass of the most massive cluster in the present sample), assuming a concentration $c_{500}=3.2$ and gas in hydrostatic equilibrium in the cluster potential with a density profile identical to that of the dark matter. }}\label{fig:Kfg}
\end{figure}

In Figure~\ref{fig:fg} we show the gas mass fraction profiles $f_{\rm gas}\, (<R) = M_{\rm gas}\, (< R)/M\, (<R)$ for the present sample. Gas masses have been calculated from the gas density profiles \citep{croston08}. Total mass profiles were calculated assuming an NFW profile with concentration $c_{500}=3.2$, the average concentration derived from the total mass profiles of the morphologically regular sample of \citet{pap05}\footnote{The dependence of concentration on total mass is negligible for the mass range we consider here \citep{pap05,buote07}.}, normalised to $\Mv$ estimated from Eqn.~\ref{eqn:Yx}. There is a clear dependence of $f_{\rm gas}$ on temperature/mass, throughout the observed temperature range, in the sense that hotter, more massive systems have higher gas mass fractions throughout the ICM. In addition, there is a clear dependence of gas mass fraction with radius in all systems, and only the most massive clusters have gas mass fractions which approach the universal value at the highest radii we are able to probe. 

In Figure~\ref{fig:Kfg} we plot the dimensionless entropy $K/K_{500}$ as a function of $f_g\, (< R) = M_{{\rm gas}} / (f_b\, \Mv)$, the fraction of a cluster's baryons in the ICM within radius $R$. $f_g$ is calculated assuming $f_b = 0.15$ ($\Omega_b h^2 = 0.022$ and $\Omega_m = 0.3$), and using total masses estimated from Eqn.~\ref{eqn:Yx}. Overplotted for comparison is an {\it unmodified} entropy distribution from the models of \citet{voit02}, derived for a cluster of $\Mv = 8 \times 10^{14}\, M_\odot$ (approximately the mass of the most massive cluster in the present sample), assuming a concentration $c_{500}=3$ and gas in hydrostatic equilibrium in the cluster potential with an  identical density profile to that of the dark matter. This particular representation makes explicit both the dependence of the entropy distribution on baryon (gas) fraction, and the mass/temperature dependence of the baryon (gas) fraction itself. The gradual translation of the profiles to the left hand side of the plot is due to a systematic lack of baryons (gas) in low temperature systems relative to high temperature systems. 

\begin{figure}[]
\begin{centering}
\includegraphics[scale=1.,angle=0,keepaspectratio,width=1.05\columnwidth]{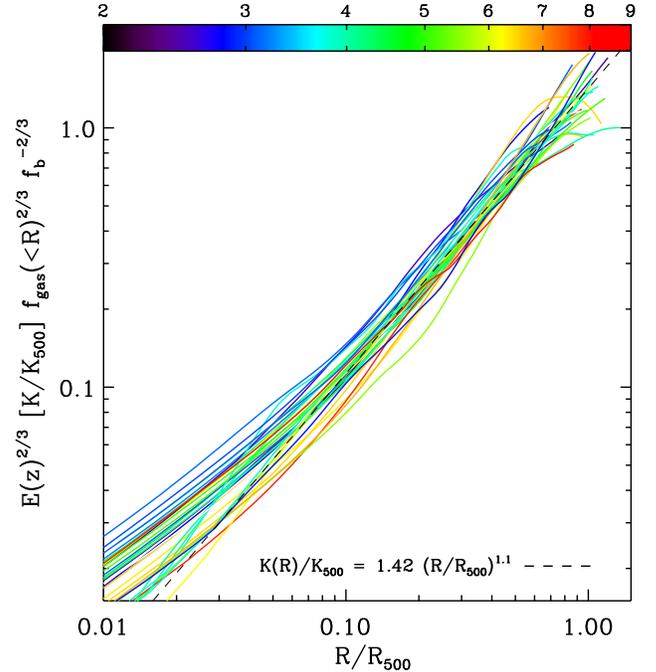}
\end{centering}
\caption{{\footnotesize Entropy profiles multiplied by the gas mass fraction profile. The dashed line is the predicted entropy distribution from the non-radiative simulations of \citet{vkb05}.}}\label{fig:KfgK500}
\end{figure}

The consequence of entropy modification is thus to remove gas (or prevent gas from accreting) in such a way as to leave both a radial {\it and} a mass dependence in the gas mass fraction. It is interesting to see whether correcting for this effect might bring the entropy profiles of our sample into agreement with the predictions from simulations. In Figure~\ref{fig:KfgK500} we show the dimensionless entropy profiles multiplied by the gas mass fraction profile $f_{\rm gas}\, (<R)$, a renormalisation that is equivalent to correcting simultaneously for the global {\it and} radial dependence of the gas mass fraction. Once renormalised in this way, the profiles are almost fully consistent, both in slope and normalisation, with the expectation from non-radiative simulations, and the dispersion drops dramatically. Slightly better agreement can be found if the simulated relation is multiplied by a factor to take into account the hydrostatic mass offset found in numerical simulations.

The above results can be explained in a number of ways. There may be a bias of gas accretion with mass, for instance due to early heating, which makes it more difficult to compress the gas into lower mass haloes. Once accreted, the gas may be removed from the hot phase by radiative cooling, which acts more efficiently in the densest central regions. However for this mechanism to be wholly responsible for the observed entropy properties would require it to affect the gas to a considerable fraction of $\Rv$ and to be preferentially efficient in low mass systems. Finally, there may be expulsion of material from the central regions towards the outskirts, perhaps via convection due to energy injection from supernovae or a central AGN, with the lowest mass systems experiencing the most central gas removal due to their shallower potential wells. 

The mass dependence of the total baryon fraction provides a way of discriminating between the competing processes. If cooling is the dominant effect, then the total baryon fraction should be almost constant across the mass range from groups to clusters as the low entropy gas is converted into stars. However, simply cooling out the low entropy gas would greatly exceed the observed mass in stars, and would lead to a galaxy luminosity function completely at odds with observations. If instead extra heating dominates, then the baryon fraction should be lower for lower mass, group-scale systems, as either early heating makes compression of gas into these haloes less efficient or AGN activity expels gas from their shallow potential wells. The observed anti-correlation in the relative dependencies of gas and stars with total mass implies that there is more mass locked in stars in systems which contain less gas \citep{lin03,gonzalez07,giodini09}.  However, recent results suggest that there is still a $\sim 3\sigma$ deficit of baryons with respect to that measured by WMAP on galaxy group scales \citep{giodini09}, implying that both cooling {\it and\/} heating must contribute to changing the thermodynamic properties of the ICM.


\subsection{Speculative scenario}

The representative nature of the \rexcess\ sample has brought to light some intriguing points outlined above, which allow us to propose a tentative scenario. It seems that about two thirds of the \rexcess\ clusters possess a significantly higher central entropy than that expected from current non-radiative cosmological simulations and consequently do not possess a cooling core. A combination of extra heating and continuous ICM mixing due to merging may have kept these systems on a higher adiabat, leading to the observed high central entropies. Some early extra heating may have occurred in the protocluster phase, which would coincide with the peak in AGN activity at $z\sim 2-3$. In this scenario the lower entropy envelope traced by the non-cool core systems (see Figure~\ref{fig:KK500}) could indicate the level of early extra heating. The distribution of central entropy above this lower envelope would then be produced by later heating and gas mixing during mergers, with the least relaxed objects having the higher central entropy, as observed. These processes will inhibit formation of a stable cool core and naturally lead to redistribution of the gas to the outskirts, acting most importantly in low mass systems, corresponding to the observed behaviour of the gas mass fraction. 

In contrast, about one third of the \rexcess\ sample possess a cool core. The clear association of the BCG with the bottom of the potential well \citep{haarsma09} and their regular X-ray morphology testifies to the relaxed nature of these objects. These systems presumably experienced a less chaotic early dynamical history leading to a modest entropy elevation due to mixing (if any), and may have undergone less early extra heating, allowing them to develop a cool core at a relatively young age. The natural reduction of entropy due to cooling while the gas is still in the hot phase, due to the combination of a temperature drop and the consequent increase in gas density needed to keep pressure balance,  may bring the profiles into line with the observed power law behavior. However, cooling must be limited by some finely-tuned feedback process to prevent a significant fraction of the gas from disappearing from the hot phase, which would  lead  to a net increase in entropy, at variance with the observations. 

Our scenario bears some resemblance to the model proposed by \citet{mccarthy08}, although with increased emphasis on merger mixing as a process for setting and maintaining entropy levels in non-cool core systems.


\section{Conclusions}

Our data represent a considerable advance over those used in most previous analyses of the entropy structure and scaling in clusters. The sample of 31 clusters spans the temperature range [2-9] keV and includes systems with a variety of entropy characteristics. The objects have all been observed to approximately the same depth with the same instrument, allowing us to probe the properties of the entropy out to significant fractions of $\Rv$ (at least $R_{1000}$ for all systems, and at least out to $\Rv$ for thirteen systems), which is essential to determine the radial extent of the effect of non-gravitational processes on the ICM. 

In the inner regions, there is a mass dependent entropy excess with respect to theoretical expectations derived from cosmological numerical simulations including only gravity. At larger radii, the mass dependence weakens and the dispersion drops dramatically. The mass dependence disappears at $\Rv$, and the entropy normalisation is, within the relatively large observational and theoretical uncertainties, in agreement with the expectations from non-radiative numerical simulations. This behaviour is mirrored in the entropy scaling relations, which are non-self similar at small radii but are compatible with self-similar at $\Rv$. While similar results were found for a sample of cool core clusters by \citet{nkv07}, it is important to note that in the group regime, at temperatures lower than 2 keV, the entropy normalisation of morphologically relaxed systems has been found to be significantly higher than predicted \citep{sun09}. 

In the inner regions there is considerable dispersion, with a distinct segregation in residuals, in the sense that cool core clusters show the least deviation and morphologically disturbed systems show the most deviation with respect to expectations from non-radiative simulations. This dependence disappears at $\sim R_{1000}$. This clear association of unrelaxed morphology and elevated central entropy would suggest either that cool cores are destroyed by mergers, or that cool cores have never been able to form in these systems. 

Fitting the entropy profiles with a power law plus constant model allows us to constrain $\alpha$, the power law slope at large radius, and also $K_0$, the central excess of entropy with respect to this power law. With the current data we cannot statistically distinguish between a bimodal distribution or a left-skewed distribution of $K_0$ in log space; however, there is certainly no evidence for strong bimodality in the present sample.
The distribution of outer entropy slopes is unimodal, with a median slope of $0.98$. Cool cores have a narrow range of outer entropy slopes (0.8-1.2) while morphologically disturbed systems have a much wider range of outer slopes (0.5-1.9), suggesting a link between the properties of the cores and the outer regions of clusters.

In seeking to explain the structural and scaling behaviour of the entropy we looked at the gas mass fraction profiles of the sample. These are strongly mass dependent. Furthermore, the gas mass fraction increases with radius in all cases. A plot of the dimensionless entropy versus baryon fraction explicitly shows the dependence of entropy on gas mass. Renormalising the dimensionless entropy profiles by the gas mass fraction profile $f_{\rm gas} (< R)$, effectively correcting simultaneously for both the mass and radial dependence of the gas mass fraction, dramatically decreases the dispersion in scaled profiles and brings them into agreement with predictions from non-radiative simulations. This provides further evidence for the underlying regularity of the cluster population, which has important implications for their use as cosmological probes.

The implication is that variations of gas content with mass and radius can explain the observed properties of the entropy distributions, and by implication, the suppression of luminosity in low mass systems \citep[e.g.,][]{pratt09}. Various physical mechanisms can impose this behaviour, but the mass and radial dependence would strongly argue for a combination of both extra energy input and radiative cooling. However, it is necessary for these mechanisms to be capable of affecting the physical properties of the gas at least up to $R_{1000}$, and perhaps beyond, in order to explain the observed offsets with respect to expectations from non-radiative numerical simulations. We discuss a tentative scenario to explain the observed behaviour of the entropy and gas mass fraction in the \rexcess\ sample, based on a combination of extra heating and merger mixing maintaining elevated central entropy levels in the majority of the population, with a smaller fraction of systems able to develop a cool core. 

In the near future, our ongoing \rexcess\ radio follow-up will help shed light on the relationship between merging activity and the lack of cool cores. Looking further ahead, observations of a similarly selected sample of galaxy groups, which are more sensitive to the effects of non-gravitational processes, would help to establish the magnitude of their impact at low masses. Precise, spatially-resolved measurements of the entropy at large radius ($R > \Rv$) will be a further essential test of theoretical models. Furthermore, a full census of the baryonic matter across the entire mass range for a {\it representative} sample of groups and clusters is essential to determine the relative contribution of the different processes in play.

\begin{acknowledgements}
We thank D. Nagai, A.J.R. Sanderson, M. Sun and A. Vikhlinin for helpful comments on the manuscript, and J. Ballet for useful discussions on statistical analysis. The present work is based on observations obtained with {\it XMM-Newton}, an ESA science mission with instruments and contributions directly funded by ESA Member States and the USA (NASA). GWP and HB acknowledge partial support from the DfG Transregio project TR33, "Dark Universe".
\end{acknowledgements}

\bibliographystyle{aa} 
\bibliography{/Users/gwp/Work/LaTeX/gwpbib}

\end{document}